\documentclass{aa}
\usepackage[utf8]{inputenc}
\usepackage{graphicx}
\usepackage{txfonts}
\usepackage{lipsum}
\usepackage{subcaption}

\usepackage{lscape}

\usepackage{placeins}

\usepackage{hyperref}
\usepackage{graphicx}
\usepackage{tikz}
\usepackage{listings}
\usepackage{amsmath,amssymb}
\usepackage{booktabs}
\usepackage{comment}
\usepackage{threeparttable}
\usepackage{tikz}
\usepackage{orcidlink}

\usepackage{hyperref}
\hypersetup{
    colorlinks=true,
    linkcolor=blue,
    citecolor=blue,
    filecolor=blue,
    urlcolor=blue
}

\usepackage{scalerel}
\usepackage{tikz}
\usetikzlibrary{svg.path}

\definecolor{orcidlogocol}{HTML}{A6CE39}
\tikzset{
  orcidlogo/.pic={
    \fill[orcidlogocol] svg{M256,128c0,70.7-57.3,128-128,128C57.3,256,0,198.7,0,128C0,57.3,57.3,0,128,0C198.7,0,256,57.3,256,128z};
    \fill[white] svg{M86.3,186.2H70.9V79.1h15.4v48.4V186.2z}
                 svg{M108.9,79.1h41.6c39.6,0,57,28.3,57,53.6c0,27.5-21.5,53.6-56.8,53.6h-41.8V79.1z M124.3,172.4h24.5c34.9,0,42.9-26.5,42.9-39.7c0-21.5-13.7-39.7-43.7-39.7h-23.7V172.4z}
                 svg{M88.7,56.8c0,5.5-4.5,10.1-10.1,10.1c-5.6,0-10.1-4.6-10.1-10.1c0-5.6,4.5-10.1,10.1-10.1C84.2,46.7,88.7,51.3,88.7,56.8z};
  }
}

\newcommand\orcidicon[1]{\href{https://orcid.org/#1}{\mbox{\scalerel*{
\begin{tikzpicture}[yscale=-1,transform shape]
\pic{orcidlogo};
\end{tikzpicture}
}{|}}}}

\usepackage{hyperref}

\begin{document}
\nolinenumbers

\title{Relations between optical spectral parameters and the (far-)infrared continuum in Type~1 active galactic nuclei}

   \titlerunning{Optical spectral parameters compared to (far-)infrared continuum}
   \authorrunning{Laki\'{c}evi\'{c} et al.}

   \author{Ma\v{s}a Laki\'{c}evi\'{c}\inst{1}\thanks{\email{mlakicevic@aob.rs}} \orcidlink{0000-0002-8231-0963},
        Luka \v{C}. Popovi\'{c}\inst{1,2}\orcidlink{0000-0003-2398-7664}, Sladjana Mar\v{c}eta-Mandi\'{c}\inst{1,2}\orcidlink{0000-0002-9995-057X},\\
        Oliver Vince\inst{1,3}\orcidlink{0009-0008-5761-3701}, \and Paola Marziani\inst{4,5} \orcidlink{0000-0002-6058-4912}
        }

   \institute{Astronomical Observatory Belgrade; Volgina 7, 11060 Belgrade, Serbia
            \and Faculty of Mathematics, University of Belgrade, Studentski trg 16, 11000 Belgrade, Serbia
            \and Shanghai Astronomical Observatory, Chinese Academy of Sciences, 80 Nandan Road, Shanghai 200030, China
            \and National Institute for Astrophysics (INAF), Astronomical Observatory of Padova, IT-35122 Padova, Italy
            \and Instituto de Astrofísica de Andaluc\'{i}a (CSIC), Glorieta de Astronom\'{i}a s{/}n, ES 18008, Granada, Spain}

   \date{Received May 20, 2026}

  \abstract
   {This study combines multi-wavelength continuum data with optical spectral properties for a specific, far-infrared (FIR)-selected sample from AKARI data of Type~1 active galactic nuclei (AGNs).}
   {We explore the relation between AGN physics and concomitant star formation processes using empirical correlations of optical and multiwavelength data.}
   {We measured several spectral parameters including the full width at half maximums (FWHMs) and velocity shifts of broad and narrow optical emission lines, as well as continuum luminosities in bands ranging from the GALEX FUV to the 160 micron band of AKARI.}
   {The data sample contains 88 AGNs observed at both optical and FIR wavelengths. The sample encompasses $\sim$50\% of Seyfert 1s with detection of both broad H$\beta$ and H$\alpha$ lines; almost all of the remaining objects can be classified as S1.9 objects where only broad H$\alpha$ is visible. The FWHM of the broad H$\beta$ and H$\alpha$ lines are anti-correlated with luminosity in most of the bands. Shifts of narrow H$\beta$, H$\alpha$, and [O\,III]5007 lines and FWHM[O\,III]5007 are correlated with optical and IR luminosities. The luminosities at 65 and 90 $\mu$m are more strongly connected with optical and other spectral properties than the ones at 140 and 160 $\mu$m. }
   {The trends presented in this paper extend the well-known Eigenvector 1 (EV1) correlations into the MIR and FIR domain. The anti-correlation trend between FWHM of broad lines and NIR and MIR luminosities is probably caused by the inclination of the torus and accretion disc: higher inclinations see less radiation of the AGN. However, the majority of objects fall into Population A along the quasar EV1 main sequence, for which narrower broad H$\beta$ profiles and FIR excess emission are commonly linked to higher Eddington ratios.}


   \maketitle

\section{Introduction}\label{sec:intro}

Understanding the connection between active galactic nucleus (AGN) spectral properties in different spectral ranges is essential for unraveling the AGN complex structure and governing processes in emission regions. The spectra of Type~1 AGNs are distinguished by the presence of both broad and narrow emission lines, which provide insights into the dynamics of the gas in the vicinity of the supermassive black hole (SMBH) at their centres \citep{Osterbrock06}. According to the standard model \citep{Urry95}, around the black hole lies an accretion disc, which is surrounded by the broad-line region (BLR) and, farther out, by the narrow-line region (NLR). These regions are enclosed by the infrared (IR)-emitting dusty torus extending out to $\sim$2-10 pc \citep{Tristam13,Ameyda20}.

Infrared observations provide an insight into the AGN environment, such as the dust composition, temperature, and physical structure, and the contribution of star formation (SF) and the AGN's energetic processes. Mid-infrared (MIR) radiation covers $\sim$5-25 $\mu$m range, while far-infrared (FIR) includes wavelengths from 25-300 $\mu$m and near-infrared (NIR) radiation covers wavelengths of $\sim$0.8-2.4$\mu$m \citep{Lamperti17}. The MIR emission from active and star-forming galaxies arises mostly from H\,II regions, photo-dissociation regions and hot dust heated by energetic X-ray to UV photons from AGNs \citep{Hernan11,Gonzalez13,Hernan15}. The NIR emission arises primarily from hot dust surrounding the AGN and from SF in the host galaxy, with a possible contribution from jets \citep{Fazeli19,Dumont19}. Attempts to connect MIR and optical spectral parameters of AGNs, to better understand AGN physical characteristics, have been made in many studies \citep{Sani10,Lacy13,Lakicevic2017,Lakicevic2018,Feltre23}. In \citet{Lakicevic2022} positive correlations of the full width at half maximum of the broad H$\beta$ line (FWHMBH$\beta$), and optical and MIR luminosities were discussed in relation to the model of dusty cone. Strong correlations have been found between optical and MIR starburst diagnostics \citep{Veilleux02,LaMassa12}. \citet{Lakicevic2018} investigated narrow- and broad-line Seyfert 1 galaxies at MIR and optical wavelengths and found that some correlations of spectral parameters are different in these two types of objects, possibly due to the different viewing angles.

Far-infrared luminosity is often used as an indicator of star formation, since the FIR emission may come from starburst regions in the host galaxy, although FIR radiation can also come from the AGN-heated dust component \citep{Netzer09,Mullaney2011,Matsuoka2015,Dai18}. A large fraction of the IR emission in AGNs comes from polar outflows in the narrow-line region (NLR; at least 40 pc from both sides) from AGNs, observed at MIR continuum \citep{Zhang13,Stalevski19,Alonso21,Isbell25}. These features provide a unique signature of the AGN's geometry and orientation relative to the observer. The problem of separating the IR emission (1-1000$\mu$m) into an AGN and a star-forming host component is still extant, although various efforts have been made to estimate these components \citep{Mullaney2011,Kirkpatrick15,McKinney21}. Usually, it is assumed that AGNs contribute negligibly to FIR emission at wavelengths $\gtrsim 100,\mu$m, and that the emission at these wavelengths is dominated by star formation in the host galaxy \citep{Mullaney2011}. In galaxies with strong star formation, the host galaxy can outshine the AGN at FIR wavelengths \citep{Shao10,Magdis13}, while AGN-heated dust can dominate the IR emission at shorter wavelengths \citep{Mullaney2011}.

Previous studies suggest that both star formation and SMBH accretion may be fueled by a common cold-gas reservoir \citep{Hickox2018}, and a statistical connection between SF and AGN activity has been observed \citep{Matsuoka2015}. \citet{Feltre13} did not find evidence that the presence of an AGN affects the star formation process in the host galaxy, but rather that the two phenomena can occur simultaneously over a wide range of luminosities. \citet{Ramasawmy19} found no clear evidence that AGN activity significantly affects SF in host galaxies. A correlation between SF and AGN luminosities is found by \citet{Netzer09} and \citet{Stanley15}; the relation between AGN luminosity and FIR luminosity has been investigated by \citet{Rosario12}. It is still unclear whether SMBHs regulate the evolution of their host galaxies \citep{Dai18}.

This study explores correlations between optical parameters and FIR (and other) luminosities in Type~1 AGNs, a comparison rarely addressed in the literature. These relations probe the link between central engine activity and the dusty environment. Optical spectral properties (e.g. FWHM, line luminosities, and shifts) trace gas kinematics and may be connected to IR emission through processes such as outflows, turbulence, and dense molecular gas. We further investigate how these sources differ from other optically visible AGNs.

A subset of Type~1 AGNs exhibits strong broad H$\alpha$ emission, while the broad H$\beta$ component is significantly weaker or even absent despite clearly detected narrow lines. This behaviour is commonly interpreted as a consequence of dust reddening within the BLR or along the line of sight, since shorter-wavelength emission such as H$\beta$ is more strongly attenuated than H$\alpha$ \citep{Goodrich1989,Maiolino2001}. The effect is consistent with large Balmer decrements indicative of substantial internal extinction and is frequently observed in intermediate Seyfert subclasses (types~1.8 and~1.9; \citealt{Osterbrock1981,Maiolino2001,Kovacevic22}).

The structure of this paper is as follows. Section~\ref{sec:Data} describes the data and methods used in this study. Section~\ref{sec:res} presents the results of the correlation analysis between the optical spectral parameters and multiwavelength luminosities. Section~\ref{sec:Discu} discusses the implications of these findings in the context of AGN physics. Finally, Sect.~\ref{sec:Conclusion} summarises the main conclusions. A cosmology with $\Omega_{\rm m}=0.3$, $\Omega_\Lambda=0.7$, and $H_0=70$ km s$^{-1}$ Mpc$^{-1}$ was adopted throughout this work.

\section{Data and methods} \label{sec:Data}

\subsection{The dataset and methods}

Explored dataset is the crossmatch between AKARI{/}FIS All-Sky Survey Bright Source Catalogue Version 1.0 \citep{Yamamura10} and the optical Type~1 AGN catalogue of \citet{Liu19} (hereafter Liu19), constructed from the Sloan Digital Sky Survey Seventh Data Release (SDSS DR7). The AKARI Survey provides fluxes at the four FIR wavelengths centred at 65, 90, 140, and 160$\mu$m \citep{Yamamura10,Murikami07}. AKARI satellite (2006-2007) operated with a 68.5 cm-diameter telescope cooled down to 6 K and observed in the wavelength range 2-180 $\mu$m from a sun-synchronous polar orbit at an altitude of 700 km \citep{Murikami07}. On the other hand, the optical data from Liu19 (z < 0.35) catalogue contain measurements of numerous parameters, such as FWHMs and the wavelength of detection of broad H$\alpha$ and H$\beta$, narrow H$\alpha$, H$\beta$ and [O\,III] lines, continuum luminosity at 5100 $\AA$ (L5100), equivalent widths (EWs) of [O\,III]5007 and broad H$\beta$ lines, the Eddington ratio, and black hole mass (M$_{{\rm BH}}$). The Liu19 catalogue also provides the photometric measurements at other bands, drawn from GALEX (FUV -- 1528 $\AA$, NUV -- 2271 $\AA$), 2MASS (J -- 1.25, H -- 1.65, and K -- 2.16 $\mu$m bands), WISE (W1 -- 3.4 $\mu$m, W2 -- 4.6 $\mu$m, W3 -- 12 $\mu$m, and W4 -- 22 $\mu$m), SDSS (u -- 355, g -- 468, r -- 616, i -- 748, and z -- 893 $\AA$) and {\sl FIRST} (radio data at 20 cm). Wavelength of detections ($\lambda_{det}$) of spectral lines were converted to the line shifts using the formula
$$shift[km/s] = c (\lambda_{det}-\lambda_{0})/\lambda_{0},$$ where $\lambda_{0}$ is the reference wavelength in vacuum and $c$ is the speed of light.

Liu19 first modelled and subtracted the continuum before fitting the emission lines. Narrow emission lines were fitted with single Gaussian profiles, while additional Gaussian components were introduced for complex [O\,III] and [S\,II] profiles. The [O\,III] and [S\,II] doublets were constrained to have the same profile, with their wavelength separations fixed to the laboratory values. The narrow H$\alpha$ and [N\,II] lines were assumed to have the same profiles and redshifts as the [S\,II] doublets, or the core of [O\,III] if [S\,II] was absent or weak. The flux ratios of the [N\,II] $\lambda6583/\lambda6548$ and [O\,III] $\lambda5007/\lambda4959$ doublets were fixed to 2.96 and 2.98, respectively. The broad H$\alpha$ profile was fitted with one or two Gaussians, while up to five Gaussian components were adopted for asymmetric or irregular broad Balmer profiles. Broad H$\beta$ was initially assumed to have the same redshift and profile as broad H$\alpha$, but its parameters were allowed to vary independently when this significantly improved the fit of the H$\beta$+[O\,III] region. The FWHMs of the broad Balmer lines were measured from the resulting multi-Gaussian model profiles. The uncertainties of the derived measurements were estimated through simulations.

\begin{figure}[h!]
   \centering
   \includegraphics[width=70mm]{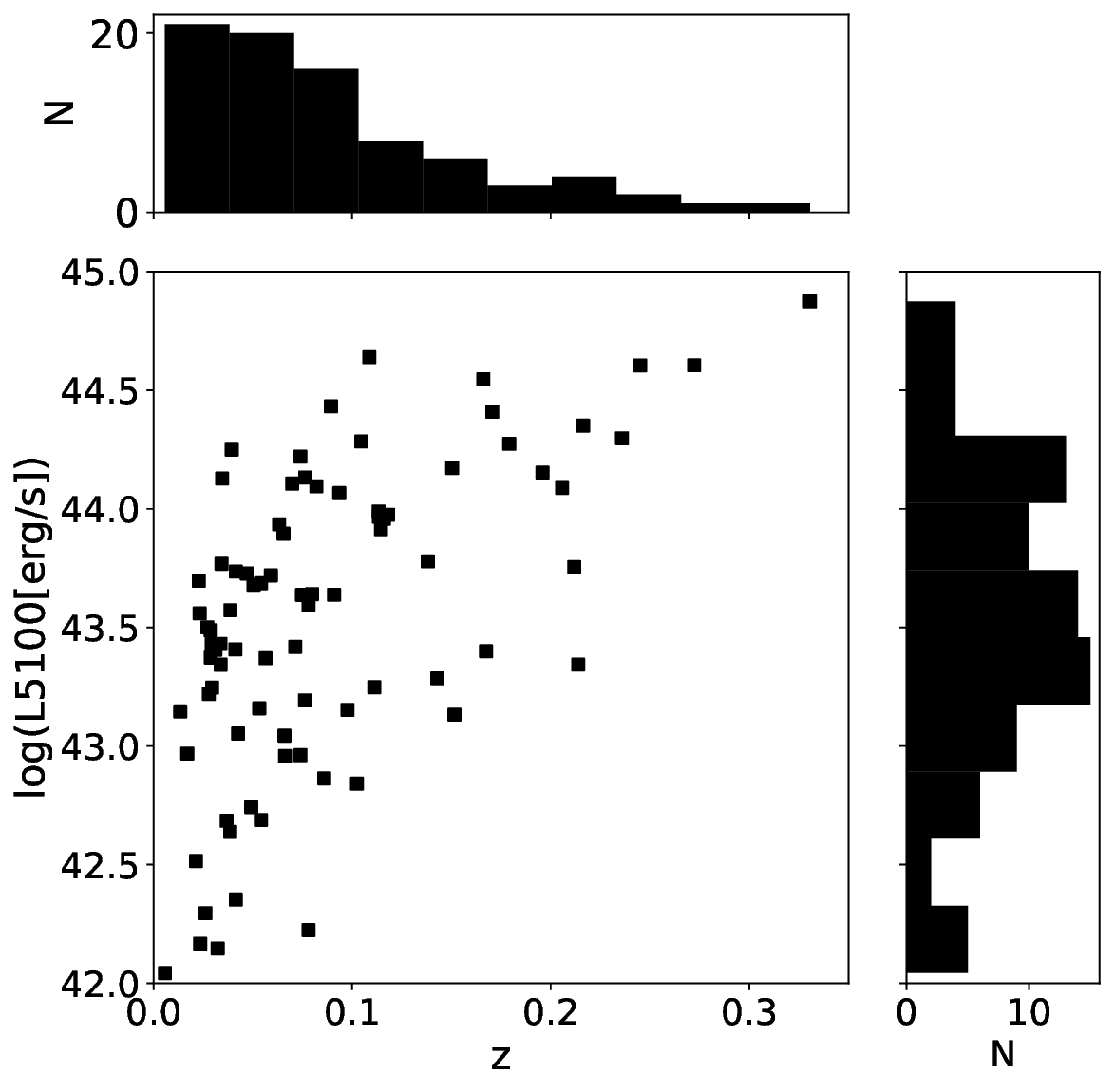}
      \caption{Dependence and distributions of L5100 and z for the chosen sample of 88 Type~1 AGNs.}
         \label{fig:fig1}
   \end{figure}

The adopted radius for above mentioned crossmatch was 7$\arcsec$, reflecting the expectation that optical emission predominantly traces the compact accretion disc of the AGN, whereas FIR emission can arise on larger spatial scales within the host galaxy (or from the dusty torus). The pointing accuracy of AKARI was $\lesssim$3$\arcsec$ \citep{Ishihara2010}, while for SDSS it is $\sim$0.1$\arcsec$ \citep{Pier03}. In that way, crossmatching Liu19 and \citet{Yamamura10} catalogues, we obtained only 88 Type~1 AGN sources, with z<0.34. The small number of these objects (just 88  from the crossmatch between the 14,000-strong Liu19 sample and $\approx$ 427,000 AKARI point sources) can lead to a conclusion that not many objects are observed both at FIR and optical wavelengths. Figure~\ref{fig:fig1} shows the dependence between redshift and L5100 and their distributions.

We computed monochromatic FIR luminosities at 65, 90, 140, and 160~$\mu$m ($\nu L_\nu$) using K-corrected flux densities and the luminosity distance. The AKARI flux densities were treated as monochromatic values at their effective wavelengths.

To assess the dependence on the assumed FIR spectral shape, we computed luminosities using two alternative prescriptions for the K-correction. First, we adopted a power-law approximation ($f_\nu \propto \nu^{-\alpha}$), whereby $\alpha$ was estimated from adjacent bands. Second, we used a modified blackbody (greybody) model with emissivity index $\beta = 1.8$, in which the dust temperature was determined individually for each source from the observed $S_{90}/S_{140}$ flux ratio using rest-frame frequencies (e.g. \citealt{Ganci2019,GendronMarsolais2026}).

The FIR luminosities derived from the two methods are highly consistent, with median offsets $\lesssim 0.1$~dex in all bands. We therefore adopted the greybody K-corrected luminosities as our fiducial values throughout the paper. Given the modest redshift range of the sample ($z < 0.34$), the K-corrections remain small and do not significantly affect the derived luminosities.

For the FIRST 20-cm peak luminosities, we assumed a standard radio power-law spectrum, $f_\nu \propto \nu^{-\alpha_{\rm r}}$ with $\alpha_{\rm r} = 0.7$, and applied the corresponding K-correction. Within our redshift range, the applied correction was modest.

We determined the source UV, optical, NIR, and MIR luminosities in the following manner. First, we corrected the photometric data for Galactic extinction using the extinction function from \citet{Cardelli89}. The mean E(B-V) values for each object were taken from IRSA\footnote{\url{https://irsa.ipac.caltech.edu}}.

In each of the four spectral domains (UV, optical, NIR, and MIR), we assumed that the spectrum could be described by a power law ($F_{\nu} \propto \nu^{-\alpha}$) and fitted the observed fluxes to determine the corresponding spectral index, \(\alpha\). This slope was then used to perform a K-correction according to $K(z) \;=\; -2.5\,\bigl(1 - \alpha\bigr)\,\log_{10}\!\bigl(1 + z\bigr)$. The K-corrected magnitudes were converted into flux densities using the appropriate effective wavelengths and zero-flux values, which are presented in the second and third columns of Table \ref{tab:luminosity}). Finally, we derived the luminosity by integrating each K-corrected flux over the filter bandpass (column 4 in Table \ref{tab:luminosity}) and multiplying by \(4\pi\,D_{L}^{2}\).

\begin{table}[ht!]
    \centering
    \begin{threeparttable}
        \caption{Filter, central wavelengths, zero-flux values $F_0$ (in $\mathrm{erg\,s^{-1}\,cm^{-2}\,\AA^{-1}}$),
            and bandwidths ($\Delta \lambda$) for each filter.}
        \label{tab:luminosity}

        \begin{tabular}{lccc}
            \toprule
            Filter
            & \(\lambda~[\mathrm{\AA}]\)
            & \(F_0~[\mathrm{erg\,s^{-1}\,cm^{-2}\,\AA^{-1}}]\)
            & \(\Delta \lambda~[\mathrm{\AA}]\) \\
            \midrule
            FUV  & 1539.0   & $1.40\times 10^{-15}$ & 442.0  \\
            NUV  & 2316.0   & $2.06\times 10^{-16}$ & 1060.0 \\
            $u$  & 3504.0   & $8.8658\times 10^{-9}$ & 634.0  \\
            $g$  & 4747.0   & $4.8307\times 10^{-9}$ & 1409.0 \\
            $r$  & 6214.0   & $2.8191\times 10^{-9}$ & 1388.0 \\
            $i$  & 7628.0   & $1.8708\times 10^{-9}$ & 1535.0 \\
            $z$  & 9068.0   & $1.3238\times 10^{-9}$ & 1409.0 \\
            $J$  & 12350.0  & $3.133\times 10^{-10}$ & 1620.0 \\
            $H$  & 16620.0  & $1.111\times 10^{-10}$ & 2510.0 \\
            $K$  & 21590.0  & $4.288\times 10^{-11}$ & 2620.0 \\
            W1   & 33526.0  & $8.180\times 10^{-12}$ & 6625.6 \\
            W2   & 46028.0  & $2.415\times 10^{-12}$ & 10423.0\\
            W3   & 115608.0 & $6.515\times 10^{-14}$ & 55069.0\\
            W4   & 220883.0 & $5.090\times 10^{-15}$ & 41013.0\\
            \bottomrule
        \end{tabular}

        \begin{tablenotes}
        \scriptsize
        \raggedright
        \setlength{\parindent}{0pt}
        \setlength{\itemsep}{-1pt}

        \item -- FUV, NUV zero fluxes: \url{https://asd.gsfc.nasa.gov/archive/galex/FAQ/counts_background.html}
        \item -- FUV, NUV effective $\lambda$ and bandwidth: \citet{Morrissey2007}
        \item -- SDSS zero flux: \url{https://www.sdss4.org/dr17/algorithms/fluxcal/}
        \item -- 2MASS: \url{https://www.ipac.caltech.edu/2mass/releases/allsky/doc/sec6_4a.html}
        \item -- WISE: \url{https://wise2.ipac.caltech.edu/docs/release/allsky/expsup/sec4_4h.html#PhotometricZP}

        \end{tablenotes}

    \end{threeparttable}
\end{table}

\subsection{The same dataset with different fitting} \label{Sec:fantasy}

To verify the correlations obtained using the Liu19 measurements and to visually inspect the spectral fitting quality, the same SDSS spectra identified by their Plate--MJD--Fiber IDs were refitted using the automatic spectral fitting code, {\sc fantasy}\footnote{https://fantasy-agn.readthedocs.io/en/latest/}. Reliable fits were obtained for 88 objects. {\sc fantasy} \citep{Ilic23} is a Python-based code designed for simultaneous multi-component fitting of AGN spectra. The code simultaneously fits the underlying broken power-law continuum together with predefined emission-line components. The Fe\,II template from \citet{Kovacevic10} is adopted. Although the fitting method differs from that used in Liu19, the derived parameters represent the same integrated spectral quantities, but their resulting values are not necessarily equivalent. In the {\sc fantasy} analysis, the Balmer profiles are explicitly decomposed into narrow-, intermediate-, and very broad-line components, and the total broad-line FWHM is measured from the combined intermediate- and very broad-line components. Thus, although the two methods measure the same spectral quantities for the same objects, differences in the adopted spectral decomposition can affect the reconstructed line profiles and, consequently, the measured line widths and shifts.

Before spectral fitting, the AGN spectra were corrected for Galactic extinction using extinction data from the NASA/IPAC Infrared Science Archive (IRSA) and the dust maps of \citet{Schlegel_1998-ApJ-500}. The spectra were then shifted to the rest frame based on redshift values provided by the SDSS database. Host galaxy contributions were removed using spectral principal component analysis, following the method of \citet{VandenBerk2001,Vanden-Berk_2006-AJ-131}, with host spectra reconstructed from a linear combination of ten quasar and five galaxy eigenspectra \citep{Yip_2004a-AJ-128-585,Yip_2004b-AJ-128-2603}; see Fig.~\ref{fig:fit28Apr4}. To preserve the integrity of narrow emission lines, these features were masked during the host subtraction process. Continuum emission was subsequently subtracted using defined windows \citep{Kuraszkiewicz_2002ApJS-143}. Final multi-component fits across the 4000-7000 $\AA$ range were performed using the {\sc fantasy} code \citep{Rakic_2022MNRAS.516.1624R}, which applies the Levenberg-Marquardt minimisation algorithm via the {\sl sherpa13 Python} package \citep{Freeman2001}. An example of the resulting fitting of the whole spectra is given in Fig.~\ref{fig:fit28Apr1}, the H$\alpha$ part of the spectra is shown in Fig.~\ref{fig:fit28Apr2}, and the H$\beta$ part of the spectra is shown in Fig.~\ref{fig:fit28Apr3}. Data obtained using the {\sc fantasy} code are provided in the \hyperlink{CDStables}{Table~\ref*{supp1}}, available in electronic form at the CDS.

\refstepcounter{table}
\label{supp1}

\section{Results} \label{sec:res}

 The present study investigates the relationships among the spectral parameters within this AGN sample. Section~\ref{domin} discusses AGN or host dominance based on the optical and FIR properties and presents the object classification. Section~\ref{sec:wr1} presents the correlations obtained using the Liu19 measurements, while Sect.~\ref{sec:wr22} examines the trends among the spectral parameters derived from the {\sc fantasy} fitting procedure. In Sect.~\ref{sec:wr5}, PCA is applied to the Liu19 data in order to identify the main parameter relationships. Finally, Sect.~\ref{fanpca} presents the PCA results based on the {\sc fantasy} spectral fitting parameters.

\subsection{The characteristics of data sample} \label{domin}

\begin{figure}[h!]
   \centering
   \includegraphics[width=88mm]{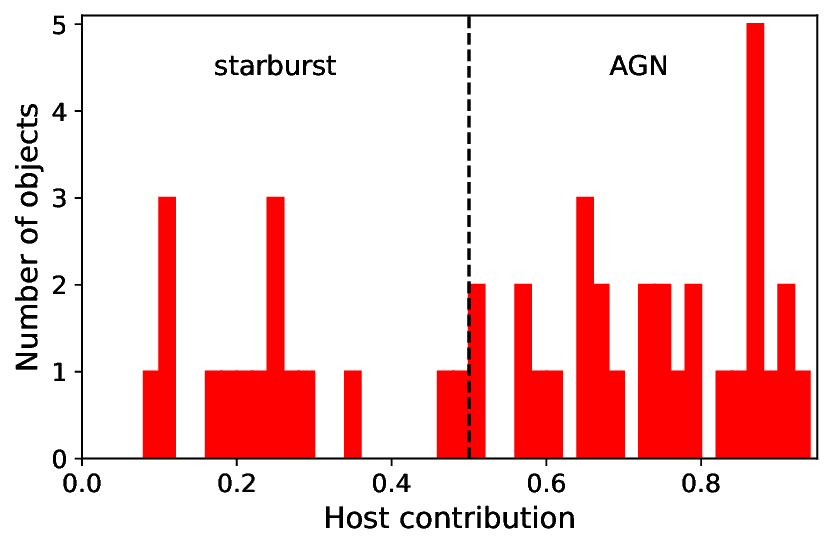}
      \caption{Distribution of host contribution in 46 objects of this sample ({\sc fantasy} measurements) that have a non-zero host contribution at optical spectra.}
         \label{fig:hco}
   \end{figure}

\begin{figure}
\centering
\includegraphics[width=88mm]{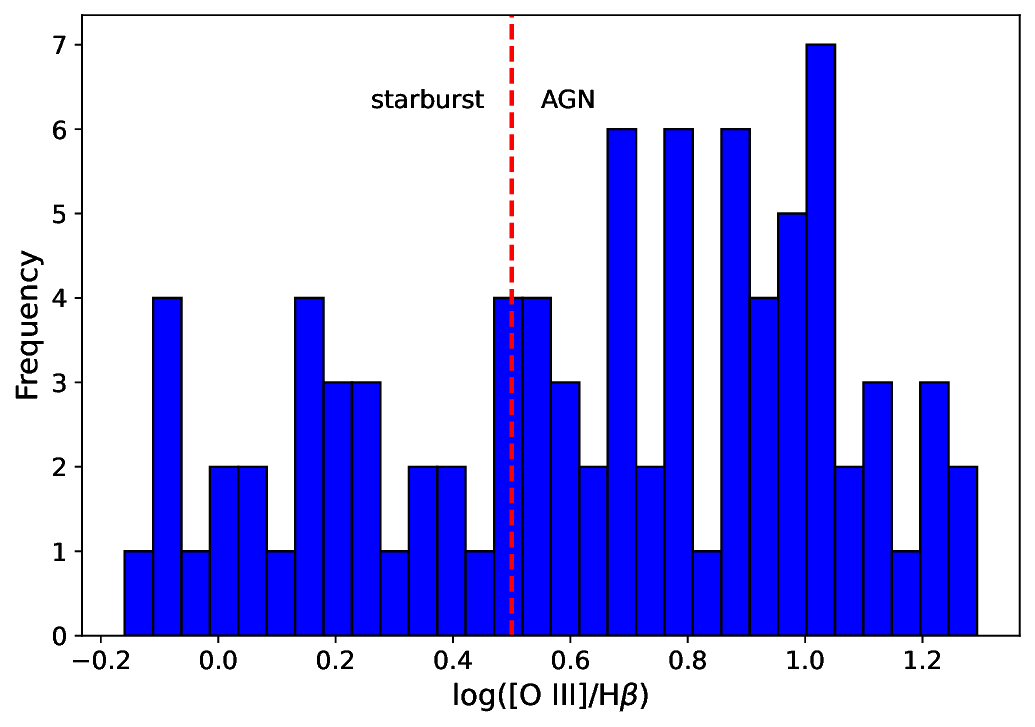}
\caption{Distribution of log([O\,III]/H$\beta$) showing that in the FIR-optical sample there are more AGN- (log([O\,III]/H$\beta$)$>$0.5) than starburst- (log([O\,III]/H$\beta$) $<$0.5) dominated objects, based on this optical ratio. \label{fig:His}}
\end{figure}

It was found that 42 of the 88 objects show no detectable host-galaxy contribution in the optical spectra. The host contribution in the wavelength interval 4160--4210~\AA\ for the remaining 46 objects is presented in Fig.~\ref{fig:hco}. Among them, 16 objects are AGN-dominated, with host contributions below 50\%, while 30 objects are host- or starburst-dominated, with host contributions exceeding 50\%. Therefore, approximately 66\% of the sample can be classified as AGN-dominated according to the optical data. A similar division was obtained using the flux ratio of the narrow [O\,III]5007 and H$\beta$ lines, R=log([O\,III]/H$\beta$), as an AGN vs. starburst diagnostic. In this case, 58 objects (66\%) are classified as AGN-dominated ($R>0.5$), while 30 objects (34\%) are classified as starburst-dominated ($R<0.5$; Fig.~\ref{fig:His}).

Objects with $\mathrm{[O\,III]/H}\beta < 3$ (26 objects) exhibit statistically significantly higher L65 luminosities compared to those with $\mathrm{[O\,III]/H}\beta > 3$ (58 objects; Mann--Whitney U test, $p = 0.012$), while no statistically significant difference is found in L90, L140, and L160 luminosities. This indicates that the observed difference is linked to enhanced emission from warmer dust components, likely associated with increased AGN heating or compact circumnuclear star formation.

\begin{figure}[h!]
   \centering
   \includegraphics[width=88mm]{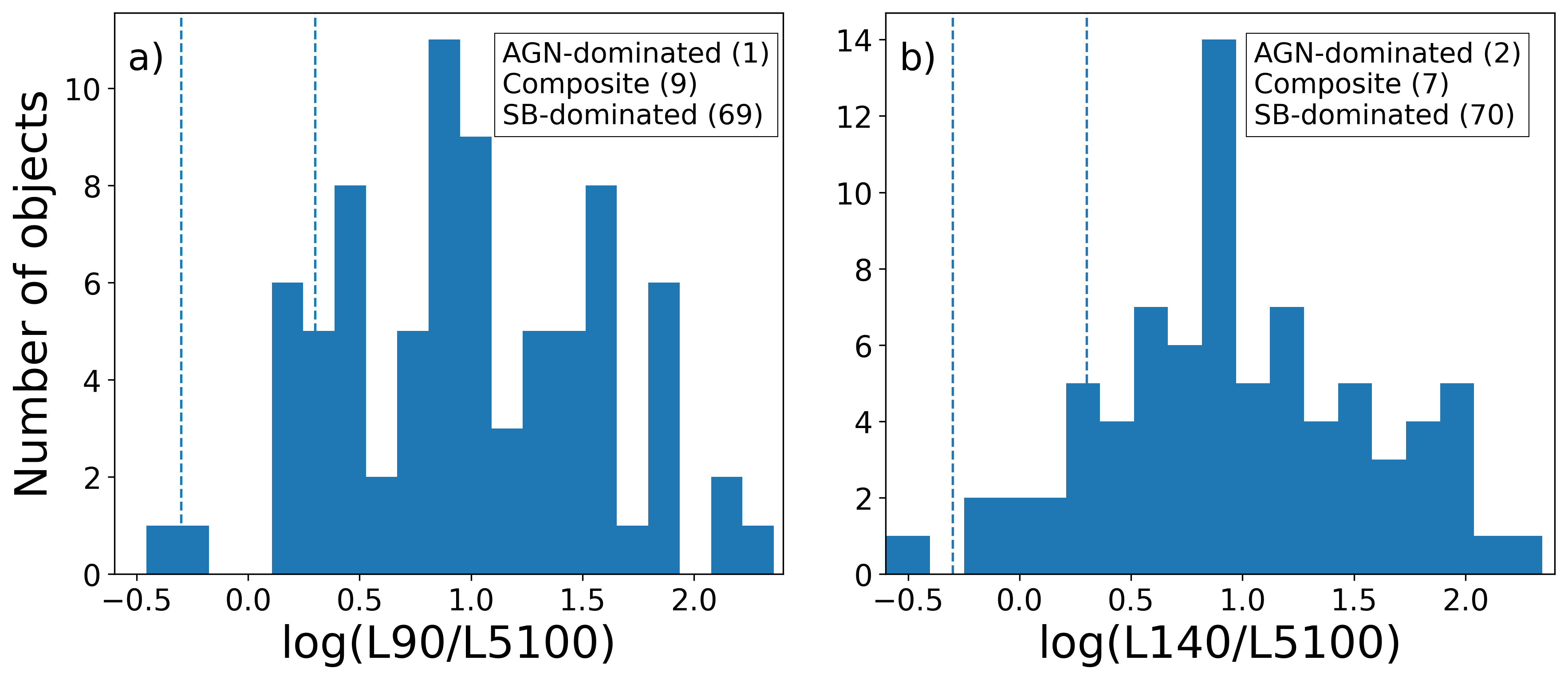}
      \caption{Distribution of log(L90/L5100) and log(L140/L5100) as indicators of AGN, starburst dominated or composite objects. The vertical dashed lines indicate the illustrative boundaries at log(LFIR/L5100) = -0.3 and 0.3 used to separate AGN-dominated, composite, and host-dominated objects.}
         \label{fig:bhb}
   \end{figure}

However, the luminosity ratios log(L90/L5100) and log(L140/L5100) can be used as approximate indicators of the relative contributions of the host galaxy and the AGN to the observed emission. Using simple illustrative thresholds for log(L140/L5100), objects with values below -$0.3$ (2 objects), between -$0.3$ and $0.3$ (7 objects), and above $0.3$ (70 objects) can be broadly classified as AGN-dominated, composite, and host-dominated systems, respectively (see Fig.~\ref{fig:bhb}a,b). Although these boundaries are arbitrary and not physically motivated, they illustrate that most objects in the sample are shifted towards high IR-to-optical luminosity ratios. This behaviour differs from the optical classification discussed above, suggesting that the sample is predominantly composite- and host-dominated in the FIR regime.

\begin{figure}[h!]
   \centering
   \includegraphics[width=88mm]{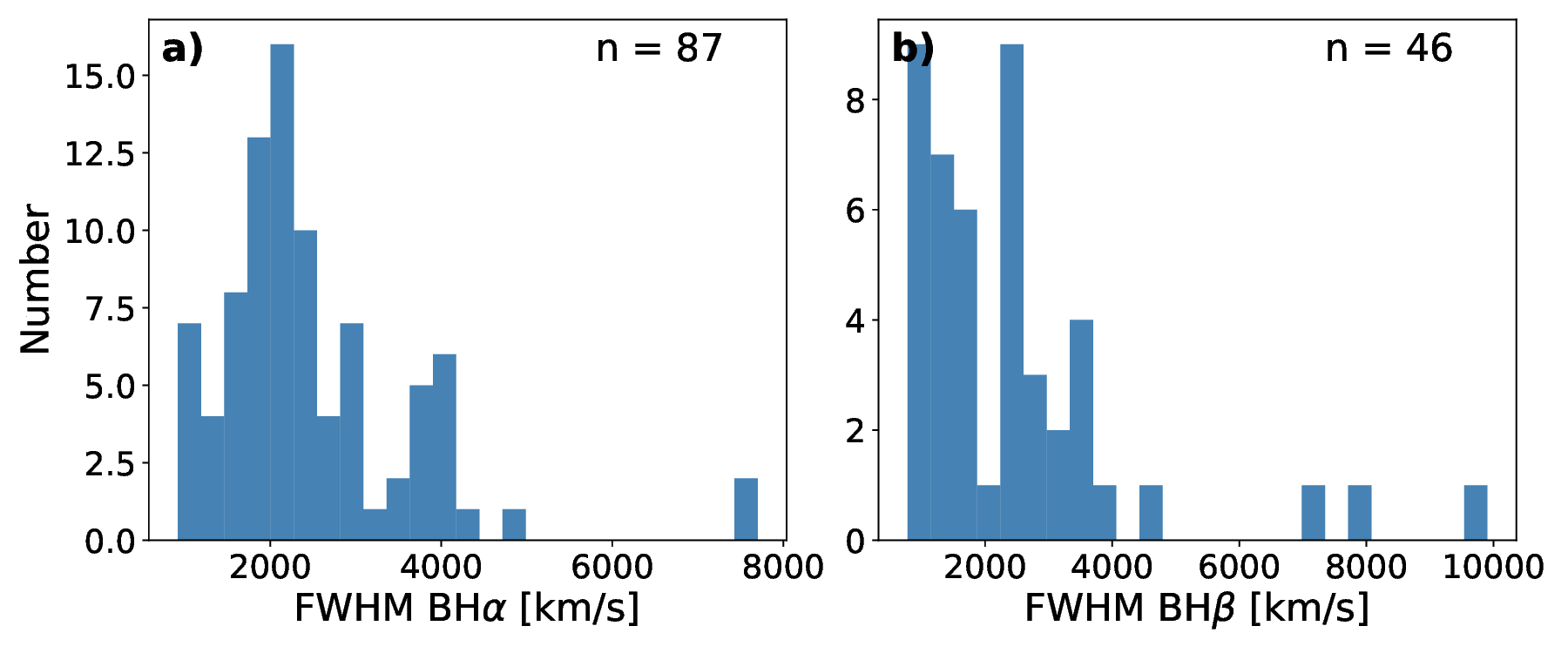}
      \caption{Distribution of FWHMBH$\alpha$ and FWHMBH$\beta$ for this sample ({\sc fantasy} measurements).}
         \label{fig:figy}
   \end{figure}

Based on the {\sc fantasy} measurements, 42 of 88 AGNs do not have broad H$\beta$, while H$\alpha$ is almost always broad, as seen in histograms in Fig.~\ref{fig:figy}a,b. There are 41 Seyfert 1.9 (46\%) objects, one Seyfert 2, 44 Seyfert 1 (50\%) objects, and two narrow-line Seyfert 1 (NLS1) objects. Adopting that radio-loud (RL) AGNs have L$_{1.4GHz}$>10$^{24}$ W Hz$^{-1}$ \citep{Meurs84}, which is 1.4 $\times$10$^{40}$ erg s$^{-1}$, we obtain five RL AGNs. Among the five RL AGNs, two exhibit narrow broad H$\beta$ profiles and two show broader H$\beta$ profiles, while in one object no broad H$\beta$ component is detected, although broad H$\alpha$ is present. One of the five objects is classified as an NLS1.

\subsection{Correlations between optical parameters and continuum luminosities from UV to radio - Liu19 data} \label{sec:wr1}

\begin{figure*}
\centering
\includegraphics[width=150mm]{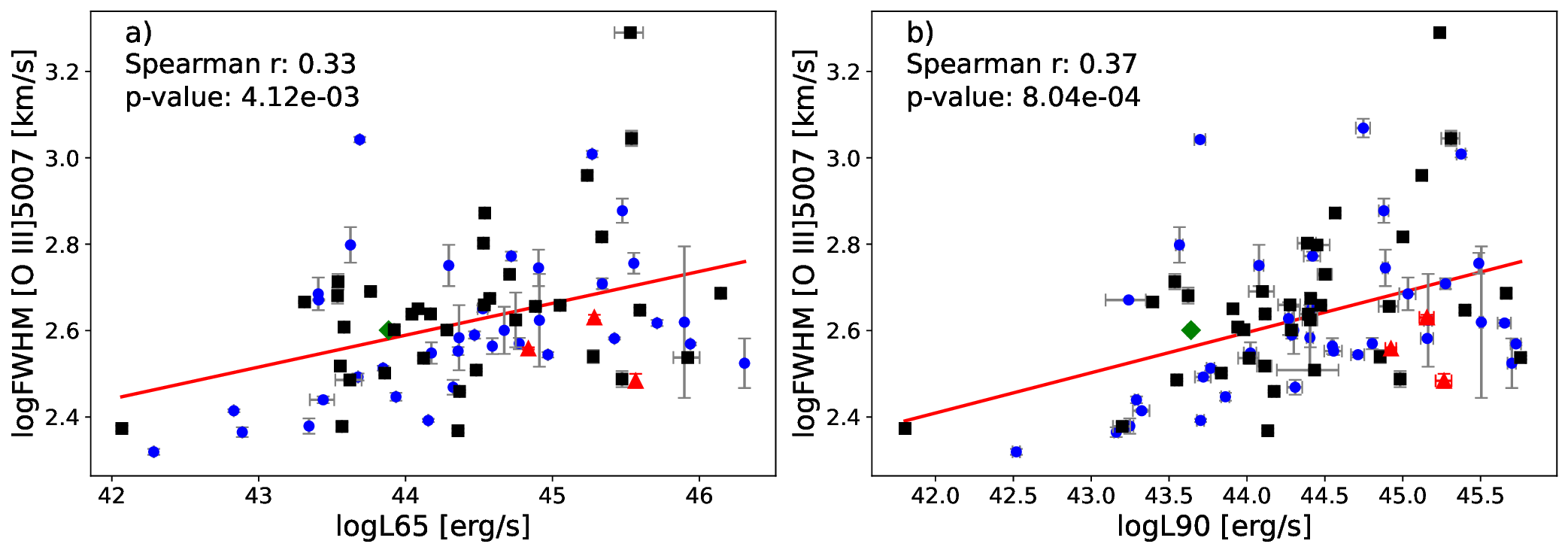}
\includegraphics[width=150mm]{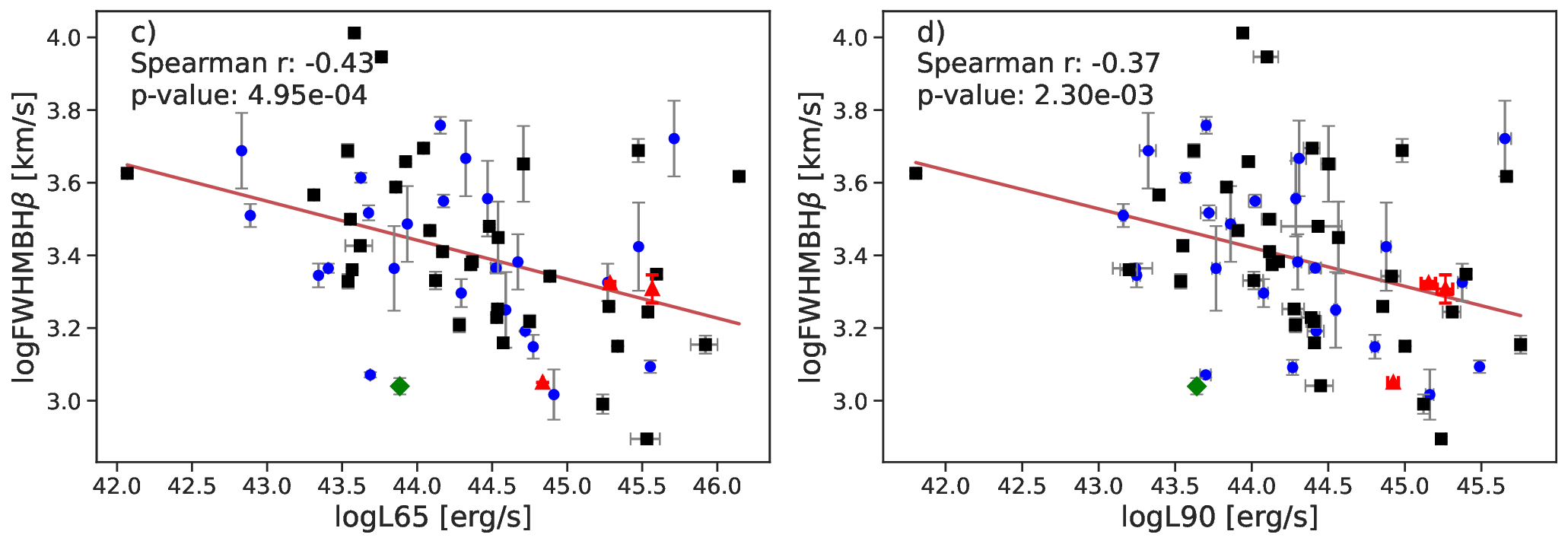}
\caption{FWHMs of narrow [O\,III]5007 and broad H$\beta$ line versus luminosities of FIR continuum at 65 and 90 $\mu$m. S1.9 objects are blue-coloured, RLs are red, NLS1s are green, and S1s are black. \label{fig:nize}}
\end{figure*}

\begin{figure*}
\centering
\includegraphics[width=150mm]{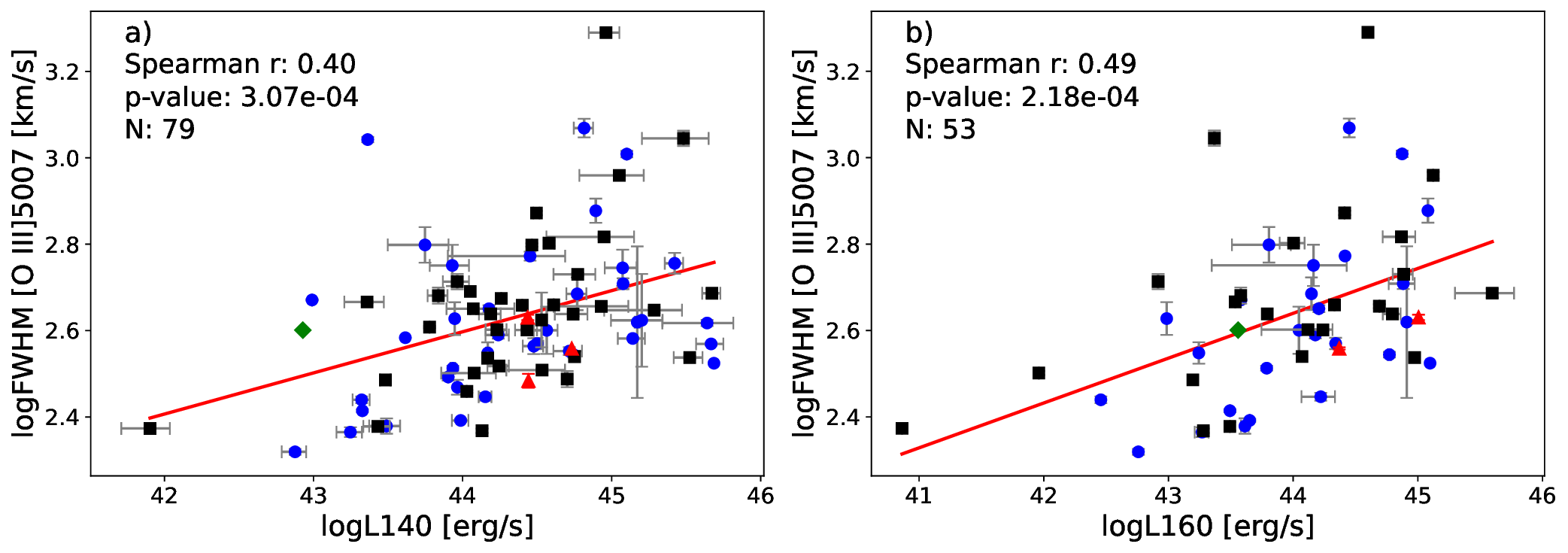}
\includegraphics[width=150mm]{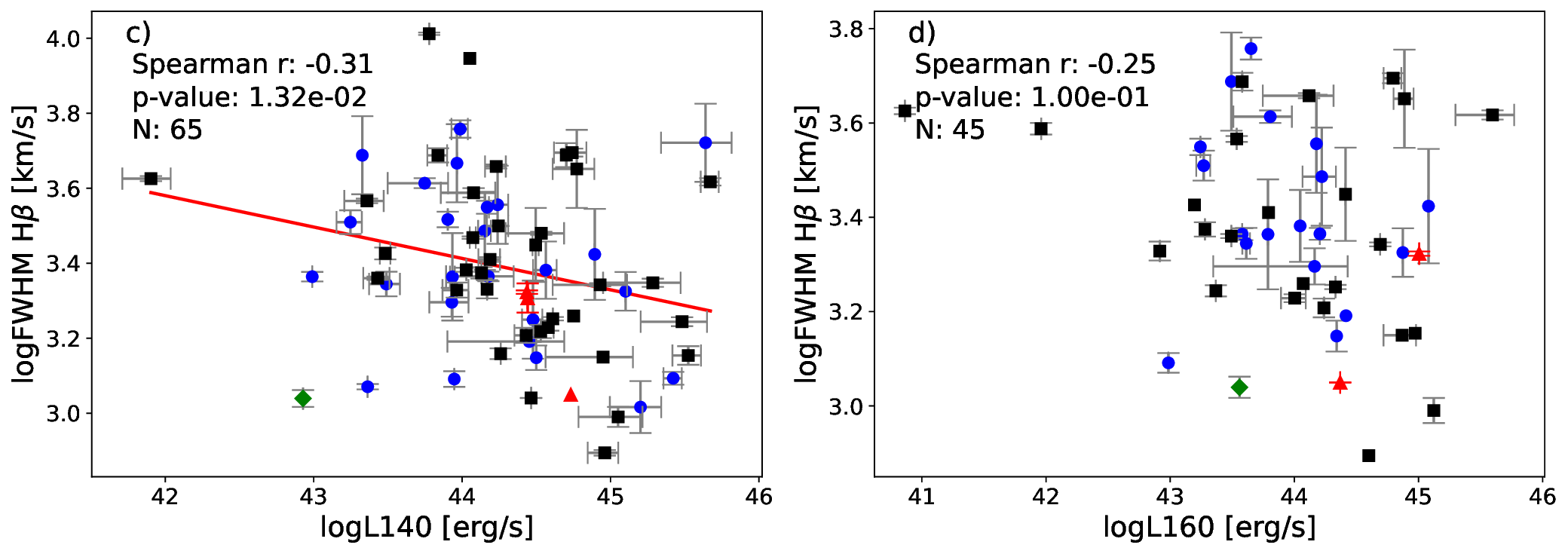}
\caption{FWHMs of narrow [O\,III]5007 and broad H$\beta$ line versus FIR continuum luminosity at 140 and 160 $\mu$m. For the symbols see the caption in Fig.~\ref{fig:nize}. \label{fig:vise}}
\end{figure*}

Although both FWHMBH$\beta$ and FWHMBH$\alpha$ were examined, FWHMBH$\beta$ was adopted in the following analysis because it exhibits stronger and more significant correlations with the investigated parameters. The complete set of correlations obtained from the Liu19 data is provided in \hyperlink{CDStables}{Table~\ref*{supp2}}, available in electronic form at the CDS, while the most relevant results are presented in Table~\ref{tab:core}. The weaker trends involving FWHMBH$\alpha$ may partly reflect the larger uncertainties introduced by blending with the [N\,II] emission lines in the automatic fitting procedure. The relations between FWHM[O\,III]5007, FWHMBH$\beta$, and the FIR continuum luminosities at 65, 90, 140, and 160~$\mu$m are presented in Figs.~\ref{fig:nize} and \ref{fig:vise}. FWHMBH$\beta$ anti-correlates with all luminosities from $i$ band luminosity to L140 (Table~\ref{tab:core} and Figs.~\ref{fig:nize},~\ref{fig:mul1}, and \ref{fig:mul2}). In contrast, FWHM[O\,III]5007 exhibits positive correlations with luminosities spanning from the $u$-band luminosity to L160, with the strongest trend observed for L160 (Table~\ref{tab:core}, Figs.~\ref{fig:nize} and \ref{fig:vise}). Throughout this work, regression lines are drawn using a standard least squares linear regression, implemented by \texttt{scipy.stats.linregress} \citep{2020SciPy}. A comparison between the regression method adopted in this work and the orthogonal regression \citep{FeigelsonBabu2012} is presented in Appendix~\ref{app:regression}.

\refstepcounter{table}
\label{supp2}

\begin{table*}
    \caption{Spearman correlations between multiwavelength luminosities and selected spectral parameters measured by Liu19.} \label{tab:core}
    \centering
    \tiny
    \begin{tabular}{|c|c|c|c|c|c|c|c|c|}
        \hline
      &FWHMBH$\beta$ &FWHM[O\,III]5007 &shiftNH$\beta$ &shift[O\,III]5007 &FWHMBH$\alpha$ & EW[O\,III]5007 &EWBH$\beta$ & Eddington ratio\\ \hline

  Lfuv         &-0.189  &0.092  &0.260  &-0.291  & -0.20 & -0.09 &0.09 & 0.22\\
                 &0.158  &0.456  &0.032  &0.016 & 0.01 &0.44 & 0.50 & 0.07\\ \hline
  Lnuv         &-0.189  &0.054  &0.250  &-0.298 & -0.151 &-0.13 & 0.1 & 0.24\\
                 &0.149  &0.648  &0.033  &0.010 & 0.203 & 0.27 & 0.46 & 0.04\\ \hline
  L5100        &-0.210&0.273  &0.237  &-0.262 & 0.086 & -0.43 & -0.23 &0.44\\
                 &0.077  &0.010  &0.026  &0.014 & 0.422 & <0.01 & 0.06 & <0.01\\ \hline
  Lu           &-0.254  &0.339  &0.319  &-0.498 & -0.276 & -0.1 & 0.22 & 0.43\\
                 &0.031  &0.001  &0.002  &0.000 & 0.009 &0.35 & 0.06 & <0.01\\ \hline
  Lg           &-0.287  &0.384  &0.354  &-0.527 & -0.254 & -0.11 & 0.15 &0.49\\
                 &0.015  &0.000  &0.001  &0.000 &0.017 & 0.30 & 0.22 & <0.01\\ \hline
  Lr           &-0.295  &0.386  &0.322  &-0.567 & -0.236 &-0.12& 0.11 & 0.51\\
                 &0.012  &0.000  &0.002  &0.000 &0.027 & 0.36 &0.35& <0.01\\ \hline
  Li           &-0.305  &0.397  &0.345  &-0.570 & -0.251 & -0.11 &0.09 & 0.52\\
                 &0.009  &0.000  &0.001  &0.000 &0.018 & 0.30 & 0.45 & <0.01\\ \hline
  Lz           &-0.313  &0.404  &0.350  &-0.572 &-0.265 & -0.14 & 0.08 & 0.55\\
                 &0.007  &0.000  &0.001  &0.000 &0.012 & 0.17 &0.50 &<0.01\\ \hline
  LJ           &-0.307  &0.436  &0.336  &-0.558 &-0.238 & -0.11 & 0.09 & 0.56\\
                 &0.010  &0.000  &0.002  &0.000 &0.029 & 0.29 &0.45 &<0.01\\ \hline
  LH           &-0.335  &0.450  &0.324  &-0.558 &-0.254 & -0.08 & 0.12 & 0.59\\
                 &0.005  &0.000  &0.003  &0.000 &0.019 &0.42 &0.32 & <0.01\\ \hline
  LKs          &-0.347  &0.448  &0.340  &-0.545 & -0.293 & -0.08 & 0.15&0.61\\
                 &0.003  &0.000  &0.001  &0.000 &0.006 &0.48 & 0.22 & <0.01\\ \hline
  LW1          &-0.325  &0.355  &0.314  &-0.494 &-0.266& -0.08 &0.16 & 0.56\\
                 &0.005  &0.001  &0.003  &0.000 &0.012 & 0.44 & 0.18 & <0.01 \\ \hline
  LW2          &-0.344  &0.372  &0.339  &-0.484 &-0.295 &-0.06 & 0.16 &0.57\\
                 &0.003  &0.000  &0.001  &0.000 &0.005 & 0.6 & 0.18 & <0.01\\ \hline
  LW3          &-0.381  &0.390  &0.368  &-0.559 & -0.317 & -0.07 & 0.08 &0.60\\
                 &0.001  &0.000  &0.000  &0.000 &0.003 & 0.50 & 0.51 & <0.01\\ \hline
  LW4          &-0.429  &0.430  &0.363  &-0.568 & -0.355 & -0.06 & 0.04  & 0.63\\
                 &0.000  &0.000  &0.001  &0.000 &0.001 & 0.55 & 0.70 & <0.01\\ \hline
 L65 & -0.426 & 0.325 & 0.298 & -0.534 & -0.309 & -0.233 & -0.151 & 0.603 \\
     & $<0.01$ & $<0.01$ & $<0.01$ & $<0.01$ & $<0.01$ & 0.043 & 0.236 & $<0.01$ \\ \hline
 L90 & -0.372 & 0.37 & 0.289 & -0.573 & -0.332 & -0.165 & -0.092 & 0.612 \\
     & $<0.01$ & $<0.01$ & $<0.01$ & $<0.01$ & $<0.01$ & 0.146 & 0.465 & $<0.01$ \\ \hline
 L140 & -0.306 & 0.396 & 0.301 & -0.511 & -0.26 & -0.114 & 0.004 & 0.55 \\
      & 0.013 & $<0.01$ & $<0.01$ & $<0.01$ & 0.021 & 0.319 & 0.975 & $<0.01$ \\ \hline
 L160 & -0.248 & 0.487 & 0.21 & -0.508 & -0.168 & -0.124 & -0.161 & 0.489 \\
      & 0.100 & $<0.01$ & 0.131 & $<0.01$ & 0.230 & 0.375 & 0.290 & $<0.01$ \\ \hline
     L20cmpeak   &-0.334  &0.239  &0.272  &-0.438 &-0.237 & -0.12 & -0.13 & 0.56\\
                 &0.009  &0.038  &0.017  &0.000 &0.039 & 0.30 &0.32 &<0.01\\ \hline
       \end{tabular}

\tablefoot{The correlations were calculated between all available luminosities and FWHMBH$\beta$, FWHM[O\,III]5007, shiftNH$\beta$, shift[O\,III]5007, FWHMBH$\alpha$, EW[O\,III]5007, EWBH$\beta$, and the Eddington ratio. The first row for each luminosity gives the Spearman correlation coefficient, and the second row gives the corresponding $p$ value. All positive data were logarithmised before calculating the correlations.
}
\end{table*}

\begin{figure*}
\centering
\includegraphics[width=170mm]{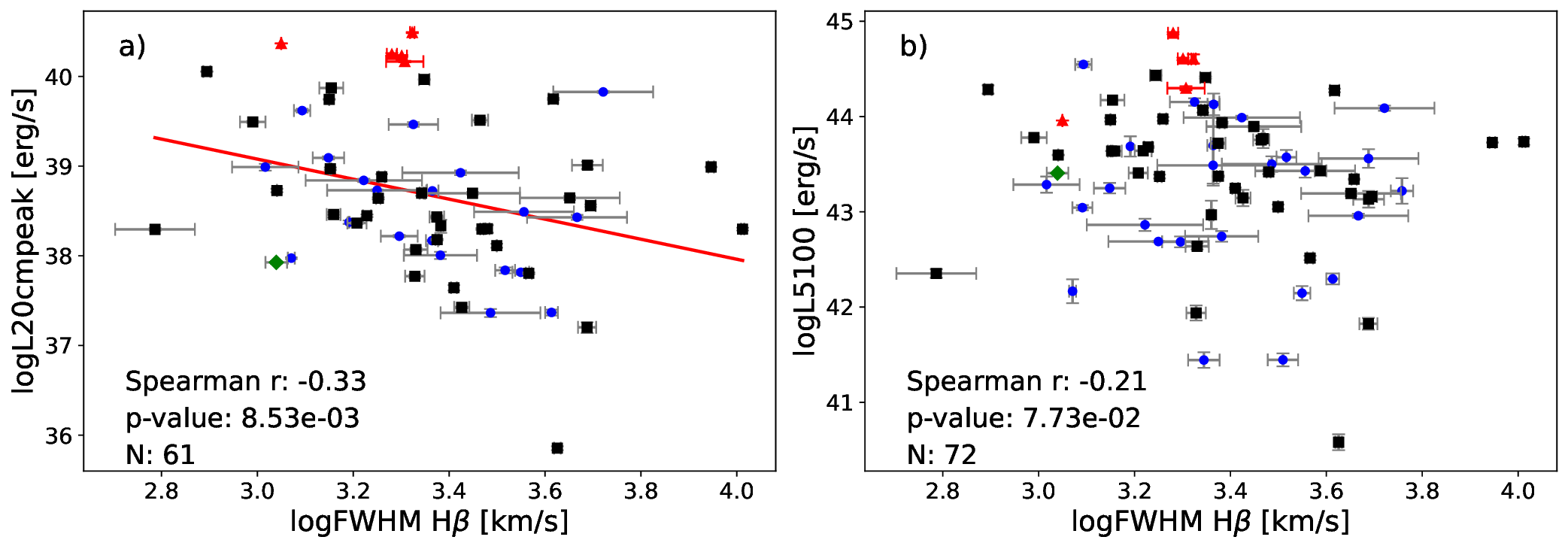}
\caption{FWHMBH$\beta$ vs radio luminosity at 20cm (a) and FWHMBH$\beta$ vs L5100 (b). For the symbols see the caption in Fig.~\ref{fig:nize}. \label{fig:rfw}.}
\end{figure*}

Figure~\ref{fig:rfw} presents the relations between FWHMBH$\beta$ and the luminosities at {\sl FIRST} radio wavelengths (20 cm) and with L5100. A weak anti-correlation is observed between FWHMBH$\beta$ and the radio luminosity, while no significant trend is found with L5100. On the other hand, FWHMBH$\alpha$ shows an anti-correlation with L5100 (Table~\ref{tab:core}). Significant positive correlations between L5100 and the luminosities in all four FIR bands ($r\sim0.6$--0.7) are presented in Fig.~\ref{fig:owt}.

\begin{figure*}
\centering
\includegraphics[width=150mm]{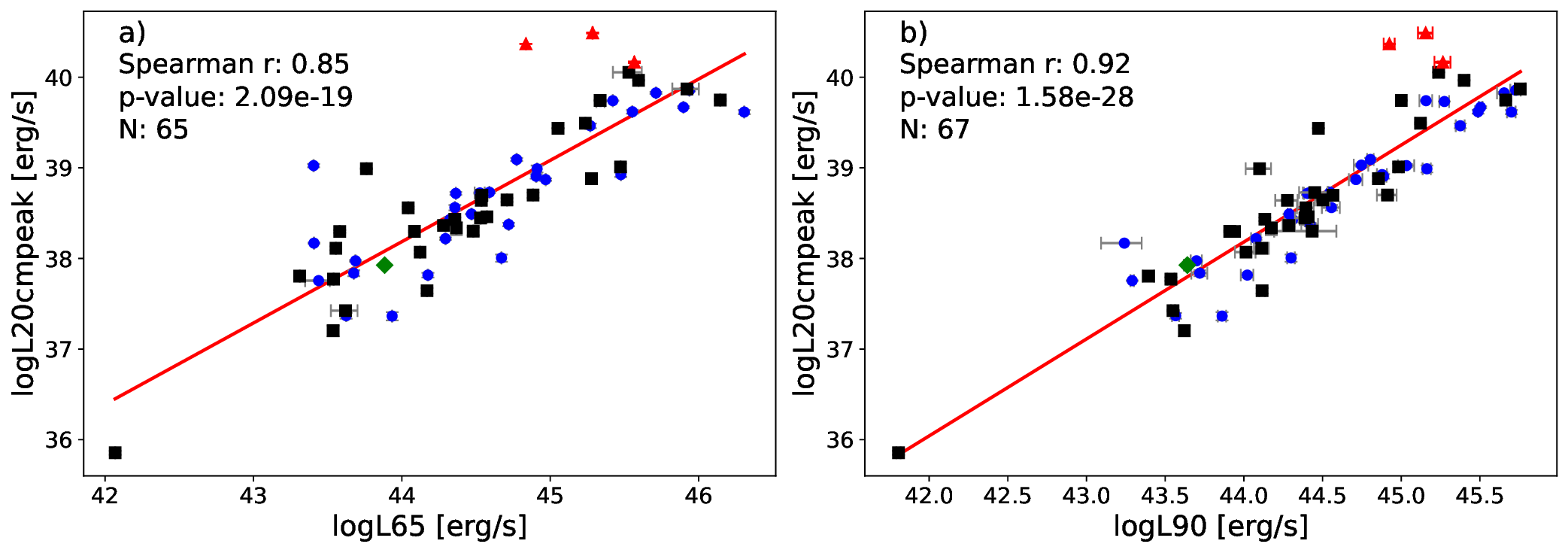}
\includegraphics[width=150mm]{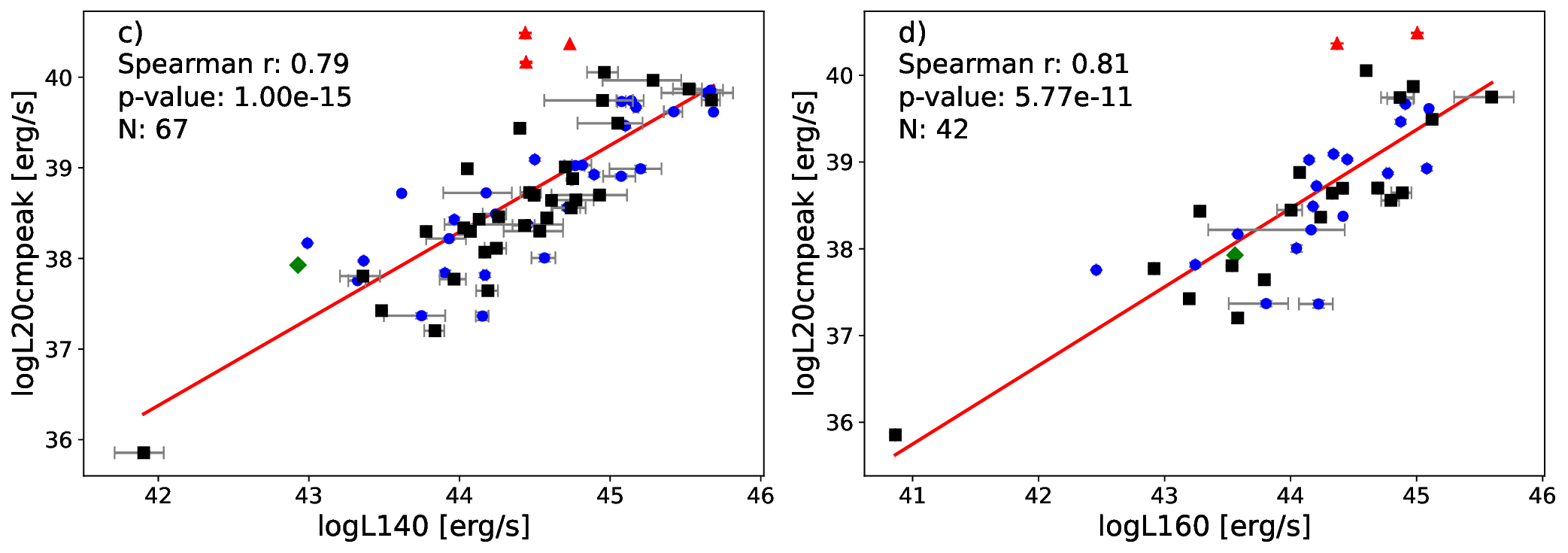}
\caption{Luminosity of radio continuum at 20cm versus FIR continuum luminosity at 65 $\mu$m (a), 90 $\mu$m (b), 140 $\mu$m (c), and 160 $\mu$m (d). For the symbols see the caption in Fig.~\ref{fig:nize}. \label{fig:radio}}
\end{figure*}

\begin{figure*}
\centering
\includegraphics[width=150mm]{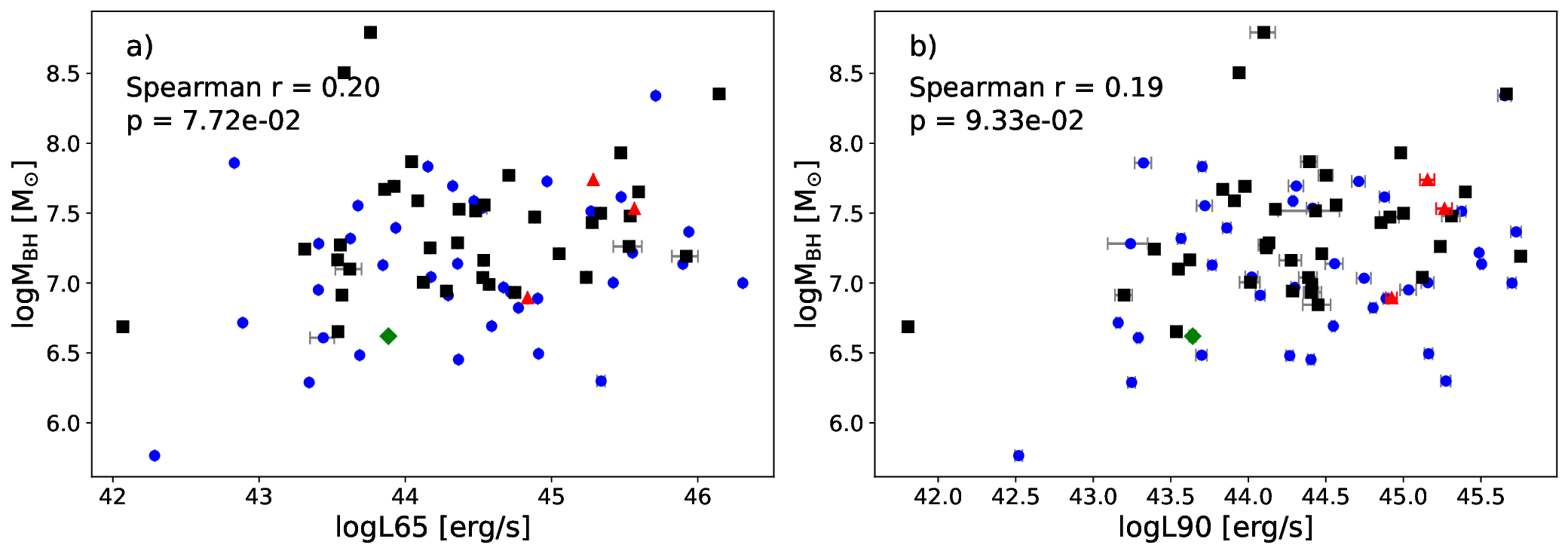}
\includegraphics[width=150mm]{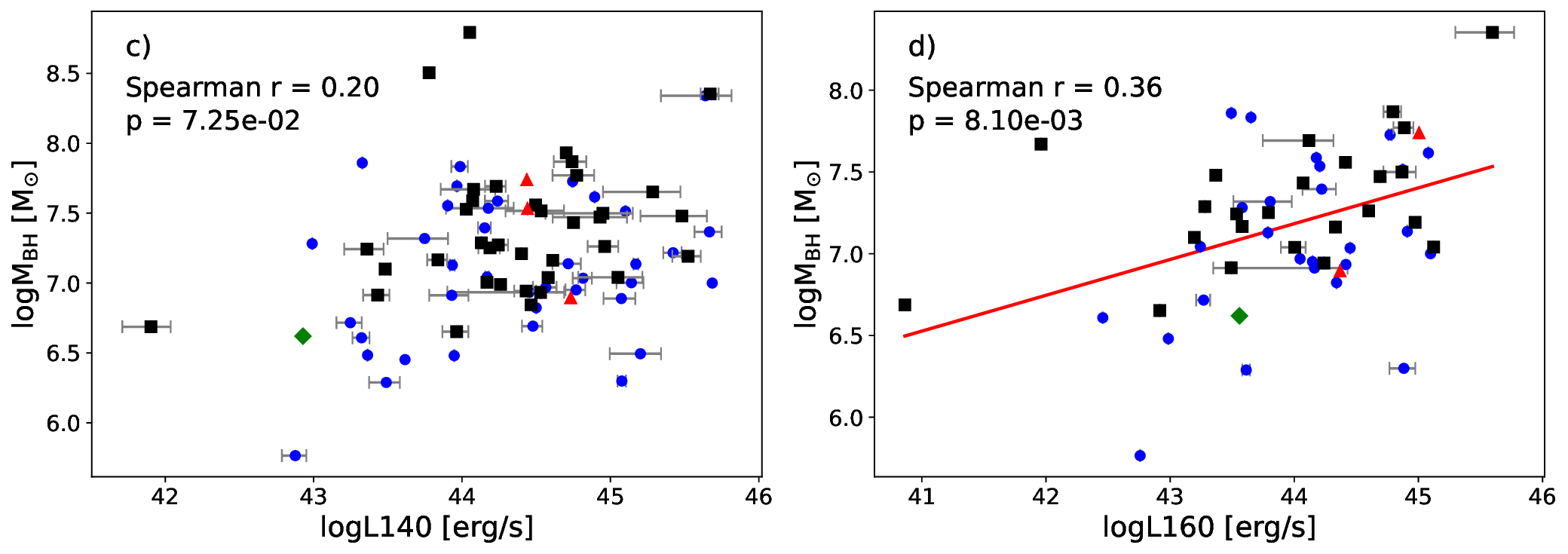}
\caption{Black hole mass (Liu19) vs FIR continuum luminosity at 65 $\mu$m (a), 90 $\mu$m (b), 140 $\mu$m (c), and 160 $\mu$m (d). For the symbols see the caption in Fig.~\ref{fig:nize}. \label{fig:mass}}
\end{figure*}

Figure~\ref{fig:radio} shows positive correlations between the radio luminosity and the FIR continuum luminosities at all four wavelengths. The relations between M$_{{\rm BH}}$ (from Liu19) and the four FIR luminosities are presented in Fig.~\ref{fig:mass}. A significant correlation is found only between M$_{{\rm BH}}$ and L160 ($r\sim0.36$). The same relations obtained using the {\sc fantasy} measurements are shown in Fig.~\ref{fig:bhbq}, where no significant correlations are detected.

\begin{figure*}
\centering
\includegraphics[width=150mm]{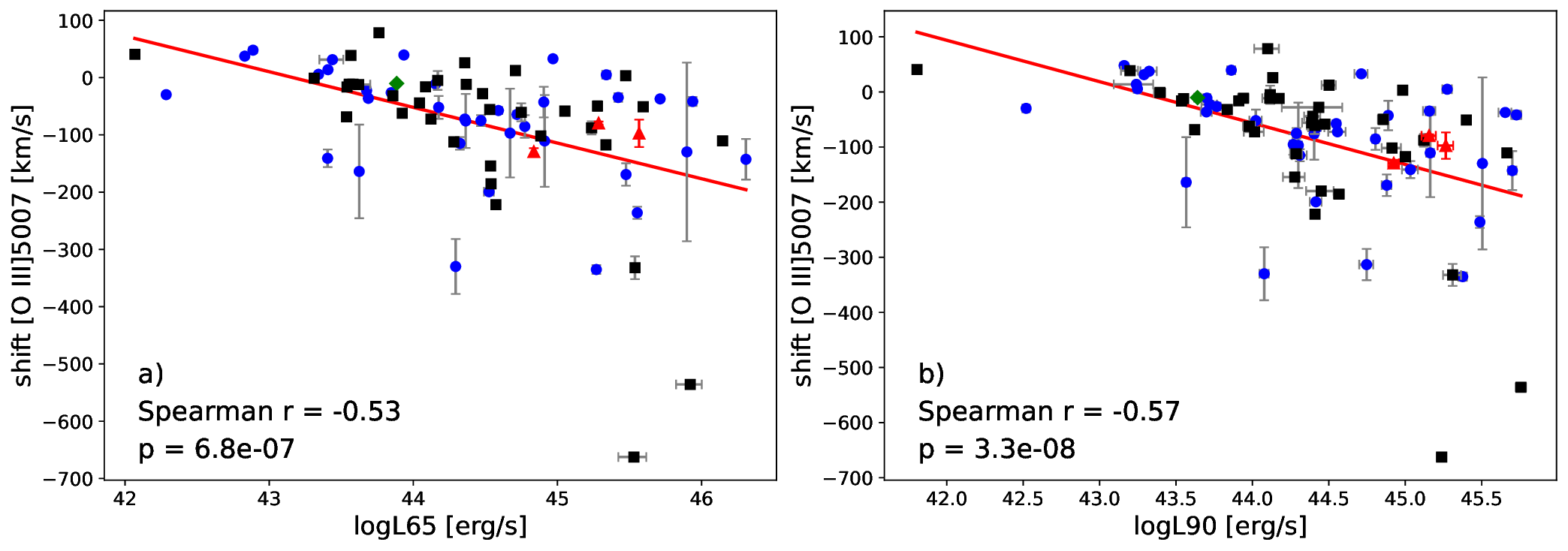}
\includegraphics[width=150mm]{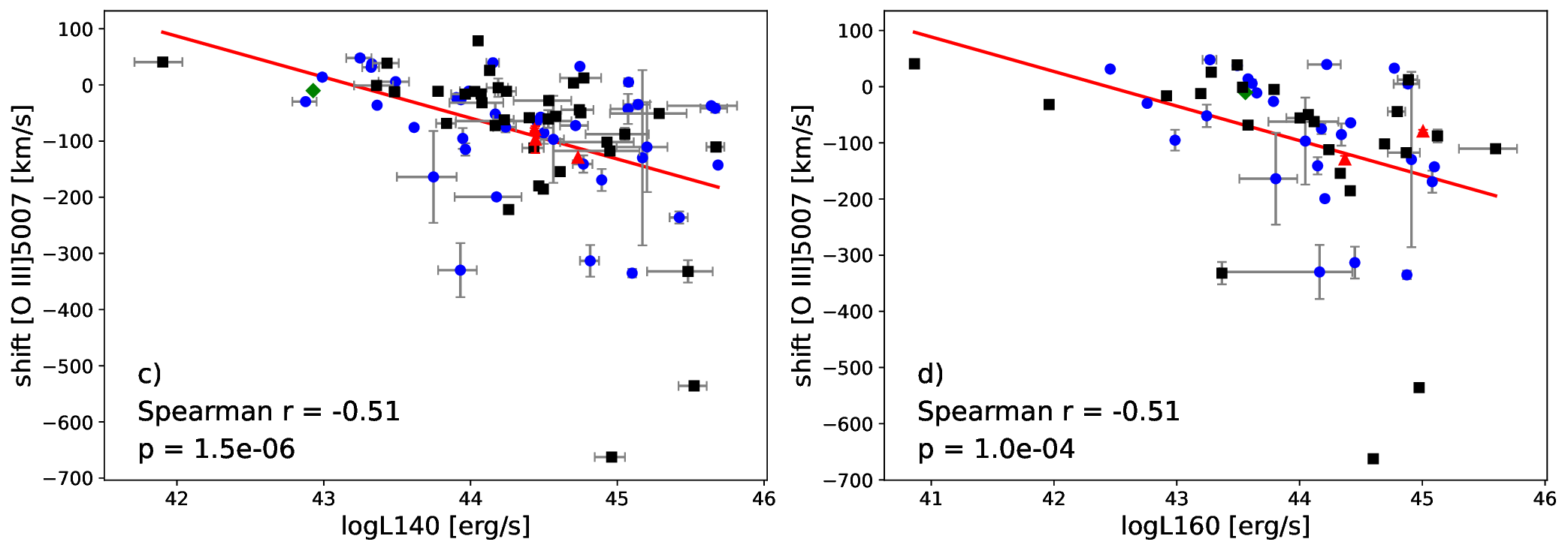}
\caption{Shift of [O\,III]5007 line vs FIR continuum luminosity at 65, 90, 140, and 160 $\mu$m in a), b), c), and d). For the symbols see the caption in Fig.~\ref{fig:nize}. \label{fig:lam1}}
\end{figure*}

\begin{figure*}
\centering
\includegraphics[width=150mm]{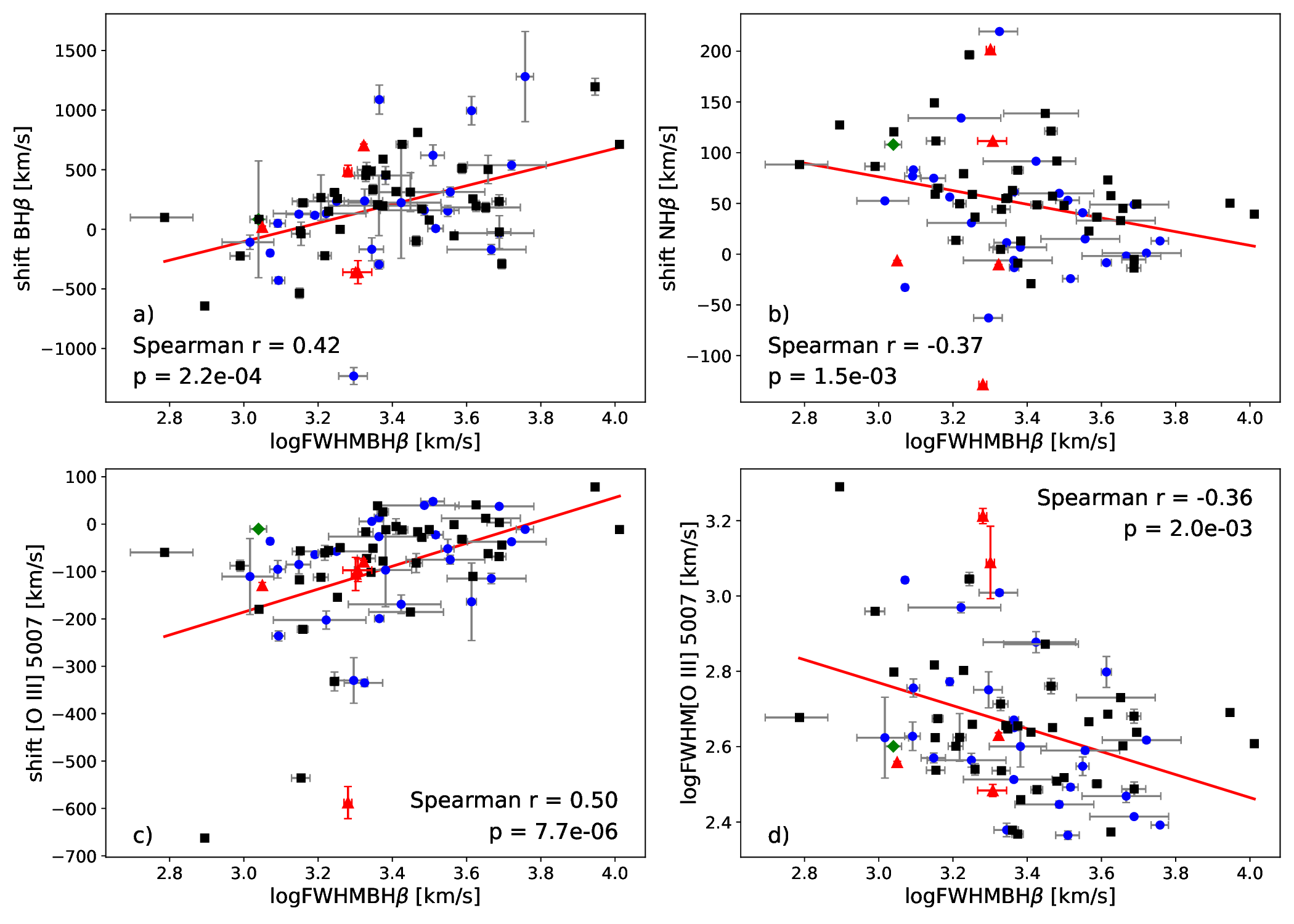}
\caption{Shifts of broad and narrow H$\beta$, [O\,III]5007, as well as FWHM[O\,III]5007 line vs FWHMBH$\beta$. For the symbols see the caption in Fig.~\ref{fig:nize}. \label{fig:lam3}}
\end{figure*}

Figure \ref{fig:lam1} (panels a-d) and Table \ref{tab:core} show significant negative correlations between the [O\,III]5007 velocity shift -- dominated by blueshifts -- and luminosity across nearly all bands. Figure \ref{fig:lam3} shows: (a, c) positive correlations between the FWHMBH$\beta$, and the velocity shifts of broad H$\beta$ and [O\,III]5007; and (b, d) negative correlations between FWHMBH$\beta$ and shiftNH$\beta$ and FWHM[O\,III]5007.

In Table~\ref{tab:core}, the Eddington ratio exhibits positive correlations with most luminosity measures; one can see in the Table~\ref{supp2} that it is anti-correlated with FWHM of broad lines and M$_{{\rm BH}}$. The EW[O\,III] shows a clear anti-correlation with L5100. Finally, FWHMBH$\alpha$ is found to be anti-correlated with L90 and with W3 and W4 luminosities; see Table~\ref{tab:core}.

\subsection{Correlations obtained using {\sc fantasy} code data} \label{sec:wr22}

The Spearman correlation matrix for all spectral parameters derived with the {\sc fantasy} fitting code, together with the corresponding $p$ values, is provided in \hyperlink{CDStables}{Table~\ref*{supp3}}, available in electronic form at the CDS. Table~\ref{tab:core1} summarises the correlations for a selected subset of parameters. Compared to the Liu19 measurements, the anti-correlation between FWHMBH$\alpha$ and optical and IR luminosities is generally stronger in the {\sc fantasy} measurements, whereas the correlations involving FWHM[O\,III]5007 and the narrow-line shifts remain qualitatively similar. Overall, the main correlation pattern is preserved, with broad Balmer-line widths being anti-correlated and FWHM[O\,III]5007 and the absolute narrow-line shifts positively correlated with most luminosities. Also it should be mentioned that EWBH$\alpha$ is correlated with FWHMBH$\beta$ (r=0.37, P=0.01).

\refstepcounter{table}
\label{supp3}

\begin{table}
    \caption{Spearman correlations between multiwavelength luminosities and selected spectral parameters measured with the {\sc fantasy} code.} \label{tab:core1}
    \centering
    \tiny
    \begin{tabular}{|c|c|c|c|c|}
        \hline
         &  FWHMBH$\alpha$ &    FWHM[O\,III]5007  &  shiftNH$\alpha$  &   shiftBH$\alpha$         \\ \hline
Lfuv & -0.28 & 0.138 & 0.16 & 0.172 \\
       & 0.022 & 0.266 & 0.193 & 0.162 \\ \hline

Lnuv & -0.241 & 0.123 & 0.192 & 0.121 \\
       & 0.041 & 0.303 & 0.103 & 0.309 \\ \hline

L5100 & -0.429 & 0.415 & 0.384 & 0.218 \\
      & 0 & 0 & 0 & 0.042 \\ \hline

Lu & -0.398 & 0.412 & 0.387 & 0.153 \\
     & 0 & 0 & 0 & 0.154 \\ \hline

Lg & -0.373 & 0.456 & 0.378 & 0.165 \\
     & 0 & 0 & 0 & 0.124 \\ \hline

Lr & -0.351 & 0.457 & 0.32 & 0.191 \\
     & 0.001 & 0 & 0.002 & 0.074 \\ \hline

Li & -0.359 & 0.461 & 0.324 & 0.227 \\
     & 0.001 & 0 & 0.002 & 0.034 \\ \hline

Lz & -0.372 & 0.46 & 0.344 & 0.217 \\
     & 0 & 0 & 0.001 & 0.043 \\ \hline

LJ & -0.342 & 0.487 & 0.288 & 0.233 \\
     & 0.002 & 0 & 0.008 & 0.033 \\ \hline

LH & -0.373 & 0.499 & 0.308 & 0.257 \\
     & 0.001 & 0 & 0.004 & 0.018 \\ \hline

LKs & -0.407 & 0.492 & 0.327 & 0.25 \\
      & 0 & 0 & 0.002 & 0.02 \\ \hline

LW1 & -0.352 & 0.396 & 0.312 & 0.147 \\
      & 0.001 & 0 & 0.003 & 0.172 \\ \hline

LW2 & -0.38 & 0.413 & 0.316 & 0.157 \\
      & 0 & 0 & 0.003 & 0.143 \\ \hline

LW3 & -0.405 & 0.444 & 0.293 & 0.229 \\
      & 0 & 0 & 0.006 & 0.032 \\ \hline

LW4 & -0.432 & 0.475 & 0.268 & 0.293 \\
      & 0 & 0 & 0.012 & 0.006 \\ \hline

L65 & -0.408 & 0.303 & 0.182 & 0.248 \\
    & 0 & 0.005 & 0.098 & 0.023 \\ \hline

L90 & -0.415 & 0.401 & 0.191 & 0.339 \\
    & 0 & 0 & 0.075 & 0.001 \\ \hline

L140 & -0.307 & 0.318 & 0.272 & 0.26 \\
    & 0.005 & 0.004 & 0.014 & 0.019 \\ \hline

L160 & -0.213 & 0.451 & 0.097 & 0.329 \\
    & 0.099 & 0 & 0.456 & 0.01 \\ \hline

L20cmpeak & -0.265 & 0.268 & 0.171 & 0.227 \\
          & 0.021 & 0.02 & 0.141 & 0.048 \\ \hline
\end{tabular}

\tablefoot{The correlations were calculated between all available luminosities and FWHMBH$\alpha$, FWHM[O\,III]5007, shiftNH$\alpha$, and shiftBH$\alpha$. The first row for each luminosity gives the Spearman correlation coefficient, and the second row gives the corresponding $p$ value. All positive data were logarithmised before calculating the correlations.
}
\end{table}

\subsection{Principal component analysis}

Principal component analysis (PCA) is a statistical technique used to reduce the dimensionality of a dataset while preserving as much variance as possible. It transforms the original variables into a new set of variables called principal components (PCs) or eigenvectors, which are linear combinations of the original variables. Eigenvectors represent the weights assigned to each original variable in forming the PCs. The proportion of variance (POV; also known as explained variance) indicates how much of the total variance in the data is captured by each PC. Principal component 1 (PC1) represents the direction in the dataset that captures the maximum variance, while PC2 captures the next highest variance, but is orthogonal to PC1. The same applies for PC3, PC4, and so on. Here we choose the set of spectral parameters from our dataset whose mutual relations we want to explore, to compare them with the relations we obtained above.

\subsubsection{PCA for Liu19 data} \label{PCAdi} \label{sec:wr5}

Table~\ref{tab:optpcaxx} presents the first four PCs obtained from the PCA applied to the selected spectral parameters of the Liu19 dataset. The corresponding POVs are reported in the last row. In this calculation 72 objects were used (where all data are available). PC1 (second column) is the most dominant in the dataset and in each column the parameters with the same sign are positively correlated, while parameters with the opposite sign are negatively correlated.

\begin{table}
    \caption{Results of PCA using optical spectral parameters from Liu19.}
    \label{tab:optpcaxx}
\centering
    \begin{tabular}{|c|c|c|c|c|}
        \hline
                  & PC1 &  PC2  & PC3 &  PC4    \\ \hline
logL90            & 0.41  & 0.24  & -0.32 & 0.12 \\ \hline
logFWHM[O\,III]5007 & 0.47  & 0.32  & 0.02  & -0.01 \\ \hline
logFWHMBH$\beta$  & -0.38 & 0.48  & 0.02  & 0.30 \\ \hline
logFWHMNH$\alpha$ & 0.29  & 0.40  & 0.72  & 0.22 \\ \hline
shift[O\,III]5007   & -0.45 & -0.18 & 0.35  & 0.37 \\ \hline
shiftNH$\beta$    & 0.28  & -0.35 & -0.22 & 0.82 \\ \hline
shiftBH$\beta$    & -0.31 & 0.55  & -0.44 & 0.17 \\ \hline
POV               & 0.41  & 0.15  & 0.13  & 0.12 \\  \hline
\end{tabular}
\end{table}

\subsubsection{PCA for {\sc fantasy} fitting data} \label{fanpca}

Table~\ref{tab:slpcaxx} presents the PCA results for a set of spectral parameters similar to those in Table~\ref{tab:optpcaxx}, obtained from the same spectra but derived using the {\sc fantasy} fitting code. The number of objects with available parameters to be used here is 87. The corresponding POVs are given in the table. The major results are similar in these two PCA analysis: L90 is anti-correlated with FWHMBH$\alpha$ and correlated with FWHM[O\,III]5007.

\begin{table}
    \caption{Results of PCA using chosen optical spectral parameters obtained using {\sc fantasy} code.}
    \label{tab:slpcaxx}
\centering
    \begin{tabular}{|c|c|c|c|c|}
\hline                       &  PC1  & PC2  & PC3 &  PC4     \\ \hline
logL90 & 0.43 & -0.12 & -0.14 & 0.33 \\ \hline
logFWHM[O\,III]5007 & 0.47 & -0.04 & -0.20 & -0.30 \\ \hline
logFWHMBH$\alpha$ & -0.40 & 0.35 & -0.13 & -0.02 \\ \hline
logFWHMNH$\alpha$ & 0.29 & 0.68 & -0.15 & 0.20 \\ \hline
shift[O\,III]5007 & 0.04 & 0.18 & 0.66 & 0.59 \\ \hline
logFWHMNH$\beta$ & 0.40 & 0.45 & 0.12 & -0.32 \\ \hline
shiftNH$\alpha$ & 0.36 & -0.32 & 0.50 & -0.22 \\ \hline
shiftBH$\alpha$ & 0.24 & -0.25 & -0.45 & 0.51 \\ \hline
POV & 0.31 & 0.17 & 0.17 & 0.11 \\ \hline
\end{tabular}
\end{table}

\section{Discussion} \label{sec:Discu}

\subsection{IR continuum luminosity compared with radio and optical continuum luminosities and M$_{{\rm BH}}$}

Radio-FIR luminosities relation for AGNs have been explored in many studies. The relation has been found to be redshift-dependent and can be used to estimate the star formation rate (SFR) from the radio luminosity \citep{Delhaize17,Wang19,Delvecchio21}, although there are claims that the observed radio emission may be connected to the AGN accretion mechanism, rather than SF \citep{Wong16}. \citet{Moric10} found that FIR/radio flux ratios are not significantly different between AGN-containing galaxies and star-forming galaxies (for z$\lesssim$0.2). A FIR--radio luminosity correlation for low-$z$ Type~1 AGN was presented by \citet{Ganci2019}, indicating that the FIR and radio power trace SF processes in sources located at the high end of the Eddington ratio distribution \citep[extreme Population~A spectral types following][]{MarzianiSulentic2014}. \citet{Delvecchio21} discovered that more massive, star-forming galaxies have a systematically lower IR-to-radio ratio, primarily due to enhanced radio emission, adding an additional layer of complexity: using radio-synchrotron emission as a proxy for SFR requires recipes dependent on baryonic mass. In the context of the present work, we just note that the maximum radio power of $\sim 1\times10^{40}\ \mathrm{erg\,s^{-1}}$ would correspond to a SFR of $\sim 400\ M_{\odot}\,\mathrm{yr^{-1}}$ \citep{Yun2001}, assuming that all radio emission originates from supernova remnants. Star formation rates of a few hundred solar masses per year are commonly found in powerful starburst galaxies. In the local Universe, the most extreme starburst-dominated systems are typically ULIRGs, with SFRs reaching up to a few $10^{3} M_{\odot},\mathrm{yr^{-1}}$ \citep{Farrah2022}. Figure~\ref{fig:radio} shows the radio luminosity at 20cm as a function of the FIR continuum luminosities. The strong correlations confirm the reliability of the cross-matching procedure and indicate a physical connection between the radio and FIR emission. The tightest correlation is obtained between radio power and L90. This is the case where (1) the contribution from the AGN torus emission should be low; (2) the wavelength roughly corresponds to the peak emission of dusty star-forming systems. The scatter increases in the cases of L140 and L160. At these wavelengths, a larger contribution from the colder host-galaxy ‘cirrus’ emission is expected, which may not be as tightly connected to the nuclear activity. Similarly, Fig.~\ref{fig:owt} presents the relation between L5100 and the FIR continuum luminosities.

Figure~\ref{fig:mass} shows the relation between M$_{{\rm BH}}$ and the FIR continuum luminosities. Among the examined bands, only L160 exhibits a statistically significant trend with M$_{{\rm BH}}$. This behaviour is consistent with the well-established connection between M$_{{\rm BH}}$ and host-galaxy properties \citep[e.g.][]{Salviander15}. No corresponding correlation is found in the {\sc fantasy} measurements (Fig.~\ref{fig:bhbq}). This study reveals an important connection between the kinematic properties of the narrow and broad emission lines -- associated with M$_{{\rm BH}}$ and ionising radiation from the AGN central engine -- and the continuum luminosities arising from the accretion disc, dusty torus, and host galaxy.

\subsection{Understanding of correlations} \label{section:tor}

The objects classified as AGN- and starburst-dominated using optical and IR diagnostics are shown in Figs.~\ref{fig:His} and \ref{fig:bhb}, respectively. Based on the optical host contribution and the emission-line ratio $R=\log([\mathrm{O\,III}]/\mathrm{H}\beta)$ (NLR), 58 objects are identified as AGN-dominated and 30 as starburst-dominated. However, the $L_{140}/L_{5100}$ ratio (Fig.~\ref{fig:bhb}) suggests that the majority of the sample may be host-dominated systems, as IR diagnostics are generally more sensitive to star formation. This is consistent with the fact that the objects are observed in the FIR.

The axis of symmetry of a dusty conus model is perpendicular to the plane of BLR; therefore, the lower inclination corresponds to a lower FWHMBH$\beta$ (FWHMBH$\alpha$). Figs.~\ref{fig:nize}, ~\ref{fig:vise}, and \ref{fig:rfw} show that the FWHM[O\,III]5007 directly depends on all FIR continuum luminosities, and that FWHMBH$\beta$ is anti-correlated with L65, L90, and radio luminosities at 20cm. This trend does not exist for colder dust (L160). Similarly, the FWHMBH$\beta$ -- L5100 correlation is absent in the Liu19 measurements (Fig.~\ref{fig:rfw}b), while a significant anti-correlation between FWHMBH$\alpha$ and L5100 is found in the {\sc fantasy} measurements (Table~\ref{tab:core1}). This difference most likely reflects the different spectral decomposition adopted by the two fitting methods. One possible explanation for the anti-correlation between broad-line FWHMs and luminosities (see Fig.~\ref{fig:mul2}) is an inclination effect. If the accretion-disc emission is anisotropic, the observed AGN continuum may decrease with increasing inclination, leading to lower observed luminosities.

In contrast to the trends found in the present sample, \citet{Lakicevic2022} reported positive correlations between FWHMBH$\beta$ and the MIR luminosities at 6 and 12~$\mu$m, as well as with L5100, for a sample of NLS1 AGNs (see their Fig.~8). \citet{Bianchi09} noticed a sharp correlation between FWHMBH$\beta$ and the luminosity of hard X-rays, for their sample of radio-quiet X-ray unobscured AGNs. An anti-correlation between FWHMBH$\beta$ and L5100 was also reported by \citet{Kaspi00}, based on a relatively small reverberation-mapped AGN sample, while in \citet{Popovic11,Kovacevic12} a positive correlation was found only for starburst-dominated objects (R=log([O\,III]/H$\beta$)) $<$ 0.5), whereas for AGN-dominated objects R=log([O\,III]/H$\beta$)$>$ 0.5) this relation was not present. In starburst-dominated AGN samples \citep{Popovic11,Lakicevic2018}, the continuum luminosity is positively correlated with the FWHM of the broad emission lines, possibly because both quantities depend on the intrinsic strength of the source, primarily related to M$_{\rm BH}$. A relation between quasar luminosity and M$_{\rm BH}$ is well established, with quasars typically spanning a range of approximately two orders of magnitude in Eddington ratio \citep{Marziani2003}. In general, larger black hole masses are associated with higher luminosities. In AGNs, where the size of the emitting region scales approximately with the square root of luminosity \citep{Bentz13} and the line broadening is dominated by virial motions, larger M$_{\rm BH}$ values are therefore expected to correspond to broader emission lines. These relations reflect the hierarchical nature of AGN systems, where larger, more massive systems exhibit higher luminosities and broader emission lines.

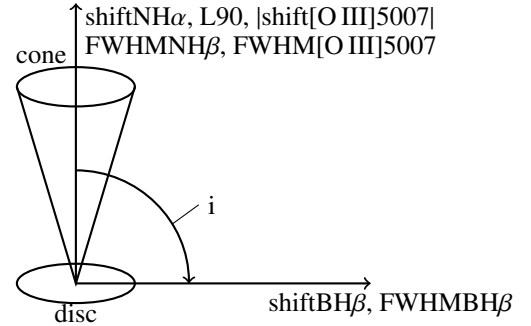
\begin{figure}[h!]
\centering
\begin{tikzpicture}[scale = 0.65]

\draw[thick] (0,4) ellipse (1.2cm and 0.4cm);
\draw[thick] (-1.2,4) -- (0,0) -- (1.2,4);

\draw[thick] (0,0) ellipse (1.2cm and 0.4cm);

\draw[->, thick] (0,0) -- (0,5.7);
\node[left] at (0,4.6) {\text{cone}};

\draw[->, thick] (0,0) -- (6,0);
\node[below] at (6.4,0) {\text{shiftBH$\beta$, FWHMBH$\beta$}};

\node at (0,-0.61) {\text{disc}};

\node[right] at (0,5.4) {\text{shiftNH$\alpha$, L90, |shift[O\,III]5007|}};
\node[right] at (0,4.8) {\text{FWHMNH$\beta$, FWHM[O\,III]5007}};

\draw[  ] (1.95,1.2) -- (2.5,1.6) node[right] {\text{i}};

\draw[->, thick] (0.,2.3) arc[start angle=90, end angle=0, radius=2.3cm];

\end{tikzpicture}
\caption{Directions where the spectral parameters are strongest: for small inclinations, high values of shiftNH$\alpha$, L90, FWHMNH$\alpha$, |shift[O\,III]5007|, and FWHM[O\,III]5007 are found, while for large inclinations, values of shiftBH$\beta$ and FWHMBH$\beta$ are larger.}
\label{fig:pic}
\end{figure}

\begin{figure}[h!]
   \centering
   \includegraphics[width=88mm]{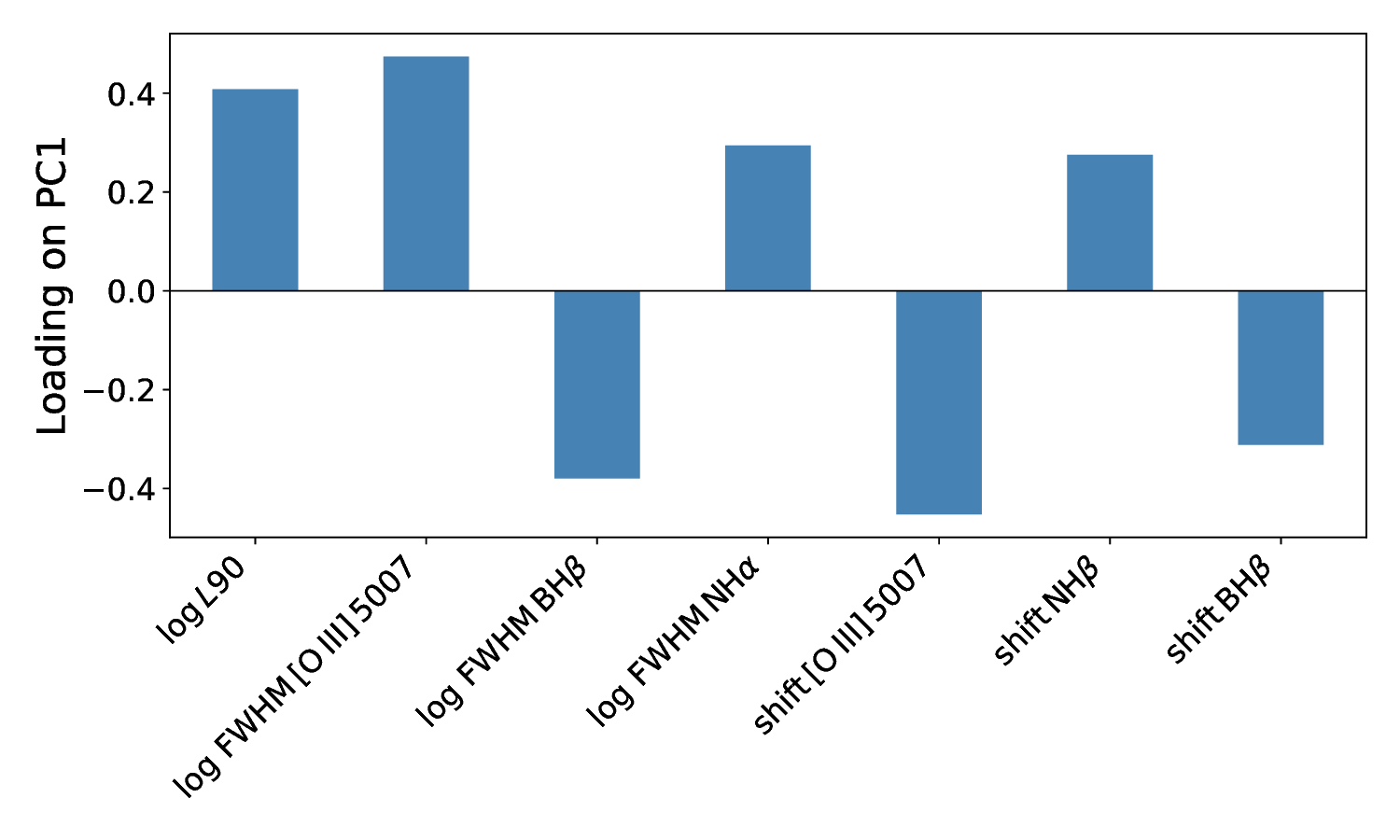}
      \caption{Loadings of spectral parameters on the first principal component (PC1). The bars show the contribution of each parameter to the PC1 eigenvector.}
         \label{fig:pic1}
   \end{figure}

In Fig.~\ref{fig:lam1}, the FIR continuum luminosities exhibit an anti-correlation with the [O\,III]5007 velocity shifts (predominantly blueshifted). Notably, the shifts of BH$\beta$ and [O\,III]5007 show a positive trend with FWHMBH$\beta$ (Fig.~\ref{fig:lam3}), whereas the shifts of NH$\beta$ and FWHM[O\,III]5007 display negative trends with FWHMBH$\beta$ (see PCA in Sect.~\ref{PCAdi}). These trends are consistent with an outflow-driven scenario, in which stronger outflows are associated with broader and more blueshifted [O\,III] emission lines \citep{Kovacevic22}. Since FWHM[O\,III]5007 traces the outflow strength, the results suggest that FIR luminosities are correlated with the outflow properties (Tables~\ref{tab:optpcaxx} and~\ref{tab:slpcaxx}). In addition, FWHMNH$\alpha$ and FWHMNH$\beta$ increase with FIR luminosity (Tables~\ref{tab:optpcaxx} and~\ref{tab:slpcaxx}). The relationships between optical spectral parameters and FIR continuum can be further visualised using the schematic model shown in Fig.~\ref{fig:pic}. For instance, at low inclination angles, the quantities shiftNH$\beta$, L90, FWHMNH$\alpha$, |shift[O\,III]5007| and FWHM[O\,III]5007 tend to be larger, while shiftBH$\beta$ and FWHMBH$\beta$ are smaller. Conversely, at higher inclinations, the former set of parameters decreases, whereas shiftBH$\beta$ and FWHMBH$\beta$ increase. The quantities |shift[O\,III]5007| and FWHM[O\,III]5007 are known to be correlated \citep{DecontoMachado2023,DecontoMachado2024}, consistent with an [O\,III] outflow oriented along the cone axis. Figure~\ref{fig:pic1} presents the loadings of the spectral parameters on the first principal component (PC1), illustrating the contribution of each parameter to the PC1 eigenvector (see Sect.~\ref{sec:wr5}). The behaviour of PC1 is broadly consistent with Eigenvector~1 (EV1), supporting the proposed link between EV1 and SF.

The [O\,III]5007 line is a well-established tracer of ionised outflows whose orientation may be linked to the geometry of the FIR-emitting structures (see Tables~\ref{tab:optpcaxx} and~\ref{tab:slpcaxx}). The observed luminosities increase towards lower inclination angles, where the FWHM of broad lines is expected to be smaller \citep{Zhang02,Afanasiev19}. This behaviour is consistent with the presence of elongated IR-emitting structures reported in the literature. Table~\ref{tab:core} shows that the strongest correlations are found between the NIR and FIR luminosities, as well as the u- and g-band luminosities, and the amplitude of the [O\,III] velocity shift. In addition, MIR emission observed by WISE, if originating in the torus, is expected to be anisotropic \citep{Stalevski2016,Netzer2016}. This is consistent with the schematic interpretation shown in Fig.~\ref{fig:pic}, which supports an orientation-based scenario.

Quasars can be broadly divided into two populations \citep{Sulentic2000}: Population~A sources, defined as having FWHMBH$\beta$ $\le 4000$~km\,s$^{-1}$, typically characterised by high Eddington ratios, strong Fe\,II emission, and H$\beta$ profiles with prominent blueshifted components and orientation-dependent broadening, whereas Population~B are sources with  FWHMBH$\beta$ $> 4000$~km\,s$^{-1}$; they generally show lower Eddington ratios, weaker Fe\,II emission, and more symmetric, Gaussian-like H$\beta$ profiles dominated by virial broadening; while redward asymmetry in the broad H$\beta$ profile is often present \citep{boroson1992,Popovic2004,Marziani2023}. Our sample mainly consists of Population A objects, as only four AGNs have FWHMBH$\beta$ larger than 4000 km s$^{-1}$. \citet{Ganci2019} show that extreme Population~A quasars (A3-A4) display radio emission consistent with SF rather than jet-dominated processes, implying a BLR and central-engine environment that differs systematically from Population~B; this provides a physically motivated basis for scenarios in which the observed FWHMBH$\beta$ need not increase with luminosity and may even show a flat or inverse trend.

The flat or even inverse FWHMBH$\beta$-luminosity trend can be physically motivated by (a) orientation effects in a flattened BLR geometry, where ‘pole-on’ views result in smaller observed FWHM \citep{Punsly2010}: quasars viewed closer to pole-on will display artificially smaller FWHM values, despite their intrinsically high luminosities; (b) radiation-driven winds, which can contribute significantly to the broad-line emission in high Eddington ratio regimes, particularly under slim-disc accretion, and can often produce blueshifted components and complex line profiles, potentially weakening the connection between the observed FWHM and luminosity \citep{Du2015, Richards2011}; (c) theoretical models that predict the occurrence of narrower emission lines at higher accretion rates \citep{Naddaf2025}; and (d) the orientation selection effect. In this work, a clear association is found between the AGN NUV/FUV and NIR/MIR/FIR luminosities, as expected if the latter are largely dominated by emission from the dusty torus. This trend is evident in the correlation matrices provided in \hyperlink{CDStables}{Tables~\ref*{supp2} and~\ref*{supp3}}, available in electronic form at the CDS.

\subsection{PCA and EV1} \label{PCAc}

The PCA results obtained using L65 are qualitatively similar to those derived for L90, whereas the analyses including L140 and L160 show weaker loadings and less pronounced trends. Therefore, L90 was adopted as the representative FIR luminosity for the subsequent PCA discussion (Table~\ref{tab:optpcaxx}; Liu19 data). The POV explained by PC1 is 41\% and it shows that FIR continuum luminosities positively correlate with FWHM[O\,III]5007, and that FIR continuum luminosities anti-correlate with FWHMBH$\beta$ as well as with shifts of BH$\beta$. |Shift[O\,III]5007| and shiftNH$\beta$ decrease with FWHMBH$\beta$ (Fig.~\ref{fig:pic}). These results confirm the relations we obtained above. A similar PCA was performed for the {\sc fantasy} measurements (Table~\ref{tab:slpcaxx}), yielding a comparable overall structure with a similar proportion of explained variance. In both analyses, the first principal component captures the anti-correlation between the broad Balmer-line widths and luminosities, as well as the positive correlation between FWHM[O\,III]5007 and luminosities. The second principal component is primarily associated with the widths and shifts of the Balmer-line components.

As shown in the PCA analysis, PC1 is aligned with an increase in FWHM[O\,III]5007, |shift[O\,III]5007|, as well as an increase in the FIR luminosities, and a decrease in FWHMBH$\beta$, FWHMH$\alpha$ (Fig.~\ref{fig:pic}). This resembles EV1 \citep{boroson1992}. \citet{Zamanov2002} demonstrated that AGN with narrower H$\beta$ profiles -- typically linked to higher $L/L_{\rm Edd}$ -- tend to exhibit larger absolute [O\,III] velocity shifts. \citet{Marziani2018m} further argued that these shifts arise from non-virial motions in the NLR, driven by disc winds and possibly amplified by orientation effects that both reduce the observed FWHMBH$\beta$ and enhance [O\,III] outflows. In addition, Fig.~\ref{fig:pic} shows that the shift of the broad H$\beta$ component increases with FWHMBH$\beta$, consistent with the trend reported by \citet{KovacevicD2017}. \citet{Panda2024} emphasised that EV1 correlations are not governed solely by the Eddington ratio, but instead arise from the combined effects of bolometric luminosity, black hole mass, and orientation.

\begin{figure}[h!]
   \centering
   \includegraphics[width=88mm]{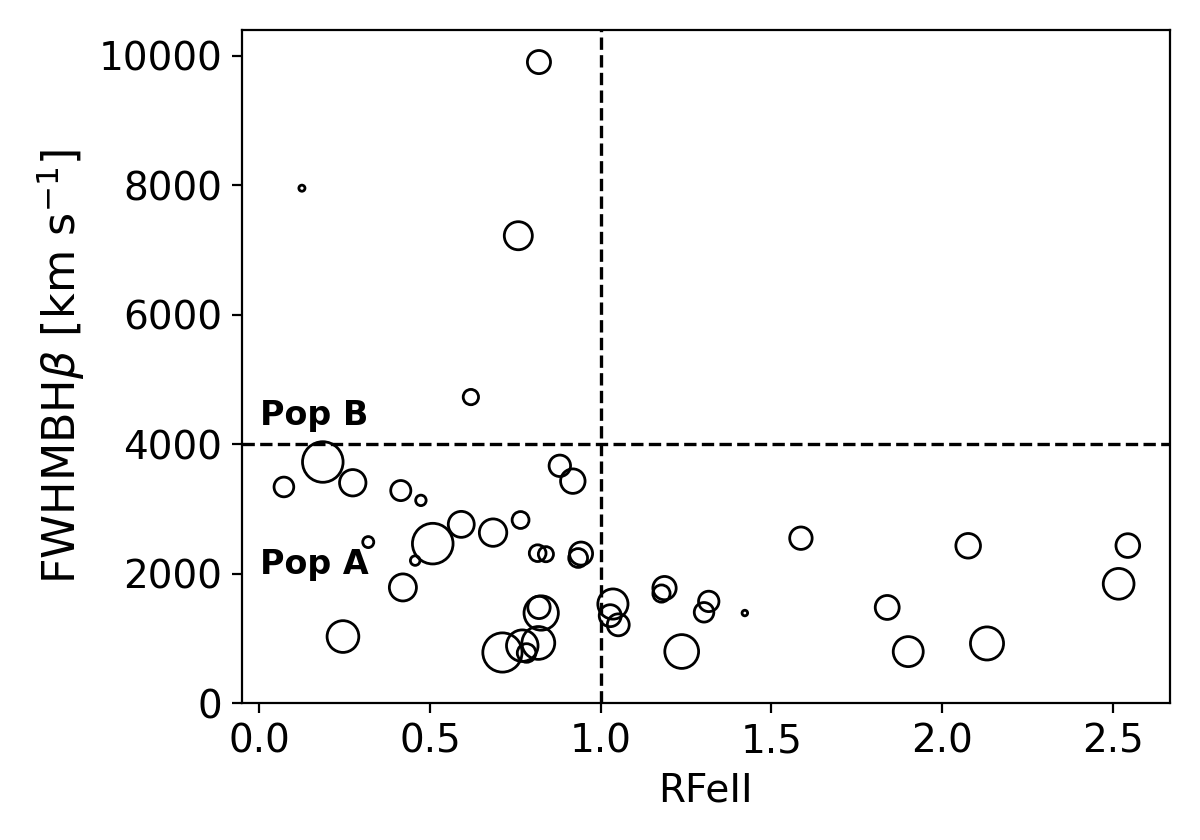}
      \caption{Relation between R${\rm Fe\,II}$ and FWHMBH$\beta$. Only the 46 AGNs that have a BH$\beta$ line are shown. The dashed lines at R${\rm Fe\,II}=1$ and ${\rm FWHMBH\beta}=4000\ \mathrm{km\ s^{-1}}$ mark the conventional boundaries within the Eigenvector~1 parameter space, separating extreme Fe\,II emitters from weaker ones, and distinguishing Population~A from Population~B quasars, respectively. The symbol sizes encode the FIR-to-optical ratio L140/L5100.}
         \label{fig:rfeii_fwhm}
   \end{figure}

The parameter R${\rm Fe\,II}$, defined as the ratio between the EWs of Fe\,II $\lambda4570$ and broad H$\beta$, is a key quantity in the EV1 context and is often linked to the accretion-disc structure and the Eddington ratio \citep{boroson1992}. Figure~\ref{fig:rfeii_fwhm} shows R${\rm Fe\,II}$ as a function of FWHMBH$\beta$, where objects with FWHMBH$\beta$ $\lesssim 4000$ km s$^{-1}$ belong to Population~A, typically associated with high-accretion sources. Only 46 Type~1 AGNs are shown in Fig.~\ref{fig:rfeii_fwhm}, since the remaining objects do not exhibit a broad H$\beta$ component. \citeauthor{Lipari1991} (\citeyear{Lipari1991}, \citeyear{Lipari1994}) showed that systems with strong Fe\,II emission tend to exhibit enhanced IR emission, suggesting a connection between Fe\,II strength and dusty environments. This connection strengthens the interpretation of R${\rm Fe\,II}$ as a tracer not only of BLR physics, but also of the connection between the accretion disc and dusty IR emission.

\subsection{Additional comments} \label{sec:host}

The observed anti-correlations between $\mathrm{FWHM}$ and luminosities may be driven by orientation effects, as the observed continuum emission from the AGN (and host galaxy) can vary with the viewing angle. It has been shown that the orientation of the BLR is not correlated with the plane of the host galaxy, indicating a largely random alignment \citep{Schmitt02,Gravity2018}. It is therefore more likely that the observed AGN luminosities decrease with increasing inclination.

However, M$_{\rm BH}$ is likely to represent a more fundamental driver of several correlations identified in this study. In particular, the relation between FWHMBH$\beta$ and shiftBH$\beta$ appears to be strongly influenced by M$_{\rm BH}$. In optically selected, starburst-dominated AGNs, the observed correlations between broad-line FWHM and luminosities are also likely governed primarily by M$_{\rm BH}$ (see Sect.~\ref{section:tor}).

A positive correlation between L160 and M$_{\rm BH}$ is found for the Liu19 dataset, whereas L65, L90, and L140 show no significant dependence on M$_{\rm BH}$ (Fig.~\ref{fig:mass}). The L160--M$_{\rm BH}$ relation may reflect the well-known connection between black hole mass and host-galaxy luminosity \citep[see][and references therein]{Salviander15}. In both datasets, the correlation between FWHM[O\,III]5007 and L160 is somewhat stronger than for other FIR luminosities (Figs.~\ref{fig:nize} and~\ref{fig:vise}), possibly indicating a link between outflow properties and the host galaxy.

\section{Conclusions} \label{sec:Conclusion}

The cross-match of the large optical (SDSS) and FIR (AKARI) datasets yields a sample of 88 Type~1 AGNs ($z<0.34$), for which key spectral parameters can be compared. The sample comprises 44 Seyfert~1 objects, 41 Seyfert~1.9s, 2 NLS1s, and 1 Seyfert~2, with 5 sources classified as RL. Approximately 48\% of the objects lack a broad H$\beta$ component, while a broad H$\alpha$ line is almost always present. Correlations were investigated by combining UV, optical, NIR, MIR, FIR, and radio luminosities with optical spectral parameters. These trends are consistent with MIR emission largely originating from the AGN dusty torus. In this context, the inclination has been proposed to be linked to FWHMBH$\beta$ \citep{Zhang02,Afanasiev19}, assuming that the symmetry axis of the dusty cone is perpendicular to the plane of the BLR. The main conclusions are summarised as follows:

\begin{itemize}
 \item For this AGN sample: optical/NIR/MIR/FIR/radio continuum luminosities are anti-correlated with FWHMBH$\beta$ and/or FWHMBH$\alpha$ (Figs.~\ref{fig:nize}c,d,~\ref{fig:rfw},~\ref{fig:mul1},~\ref{fig:mul2}; Tables~\ref{tab:core} and \ref{tab:core1}). That is different from what is obtained in \citet{Popovic11,Lakicevic2018,Lakicevic2022} and other literature, for a starburst AGN sample, where luminosities show a direct correlation with FWHMBH$\beta$.

 \item The anti-correlations of FWHMBH$\beta$ and FWHMBH$\alpha$ versus luminosities may be consequence of the viewing angle: the higher inclination may result in less AGN radiation being observed, and the seen accretion disc velocity component is larger; the higher FWHM for lower luminosities suggests a possible orientational selection effect.

 \item Anti-correlations between the broad-line widths and luminosity are expected for Population~A sources, where narrower broad-line profiles are associated with higher Eddington ratios. In such systems, higher accretion rates produce larger luminosities, while the narrower line widths are related to BLR structure, orientation, and virial scaling \citep{Sulentic2000,Marziani2018m}.

 \item L160 show a significantly stronger trend with M$_{\rm BH}$ than luminosities at lower FIR wavelengths. Luminosities at the wavelengths of 3-22 $\mu$m show more significant trends with optical spectral parameters than luminosities at 140-160$\mu$m, which are more related to the host galaxies. The FIR luminosities at 65 and 90 $\mu$m are tightly correlated with radio emission at 20 cm, following the trend expected for star-forming galaxies and radio quiet quasars. In addition, FIR luminosities are tightly correlated with the shift and FWHM of [O\,III], and the relatively low velocities involved (90\% $<$-200 km/s) leave open the possibility of starburst-driven outflows, at least for a fraction of the sample. About one third of the sources in Fig.~\ref{fig:His} have [O\,III]/H$\beta$ $<$ 3, which is commonly considered the boundary below which the emission is attributed to H\,II-region ionisation rather than AGN photoionisation. In addition, the L$_{FIR}$/L5100 ratio distribution indicates that the majority of the sources are starburst-dominated (Fig.~\ref{fig:bhb}).

 \item L65 and L90 show positive connections with the shiftNH$\alpha$, FWHMNH$\beta$, |shift[O\,III]5007|, and FWHM[O\,III]5007, while they show negative trends with FWHMBH$\beta$, FWHMBH$\alpha$, and shiftBH$\beta$ (see Sect.~\ref{PCAdi}, Fig.~\ref{fig:pic}). These trends extend the well-known EV1 correlations into the FIR domain, suggesting an increasing role of SF in highly accreting sources.
\end{itemize}

\section*{Data availability}
\hypertarget{CDStables}{}

Tables~\ref{supp1},~\ref{supp2} and~\ref{supp3} are only available in electronic form at the CDS via anonymous ftp to cdsarc.u-strasbg.fr (130.79.128.5) or via http://cdsweb.u-strasbg.fr/cgi-bin/qcat?J/A+A/.

\begin{acknowledgements}
We thank the anonymous referee for the constructive comments and suggestions that helped improve the manuscript. This work is supported by the Ministry of Science, Technological Development and Innovations of Serbia, the number of contract is 451-03-33{/}2026-03{/}200002. O.V. is partially supported by Chinese Academy of Sciences (CAS) President’s International Fellowship Initiative (PIFI) (grant No. 1295 2024VMB0006). PM acknowledges financial support from the Spanish MCIU through project PID2022-140871NB-C21 by “ERDF A way of making Europe”, and from the Severo Ochoa grant CEX2021-515001131-S funded by MCIN{/}AEI{/}10.13039{/}501100011033. Funding for the Sloan Digital Sky Survey (SDSS) and SDSS-II has been provided by the Alfred P. Sloan Foundation, the Participating Institutions, the National Science Foundation, the U.S. Department of Energy, NASA, the Japanese Monbukagakusho, the Max Planck Society, and the Higher Education Funding Council for England. The SDSS Web Site is available online \footnote{\url{http://www.sdss.org/}}. The SDSS is managed by the Astrophysical Research Consortium for the Participating Institutions listed at the above web site. This research has made use of data from the SDSS Seventh Data Release \citep{Abazajian2009}. This research has made use of the VizieR catalogue access tool, CDS, Strasbourg, France (DOI : 10.26093{/}cds{/}vizier). The original description of the VizieR service was published in 2000, A\&AS 143, 23. This research is based on observations with AKARI, a JAXA project with the participation of ESA. Part of the analysis presented in this work was done with {\sl TOPCAT} \footnote{\url{http://www.star.bris.ac.uk/~mbt/topcat/}}, developed by
M. Taylor.
\end{acknowledgements}

\bibliography{referencesL25}

\begin{appendix}

\section{Fitting with {\sc fantasy}}

\begin{figure*}[h!]
\centering
\includegraphics[width=0.9\textwidth]{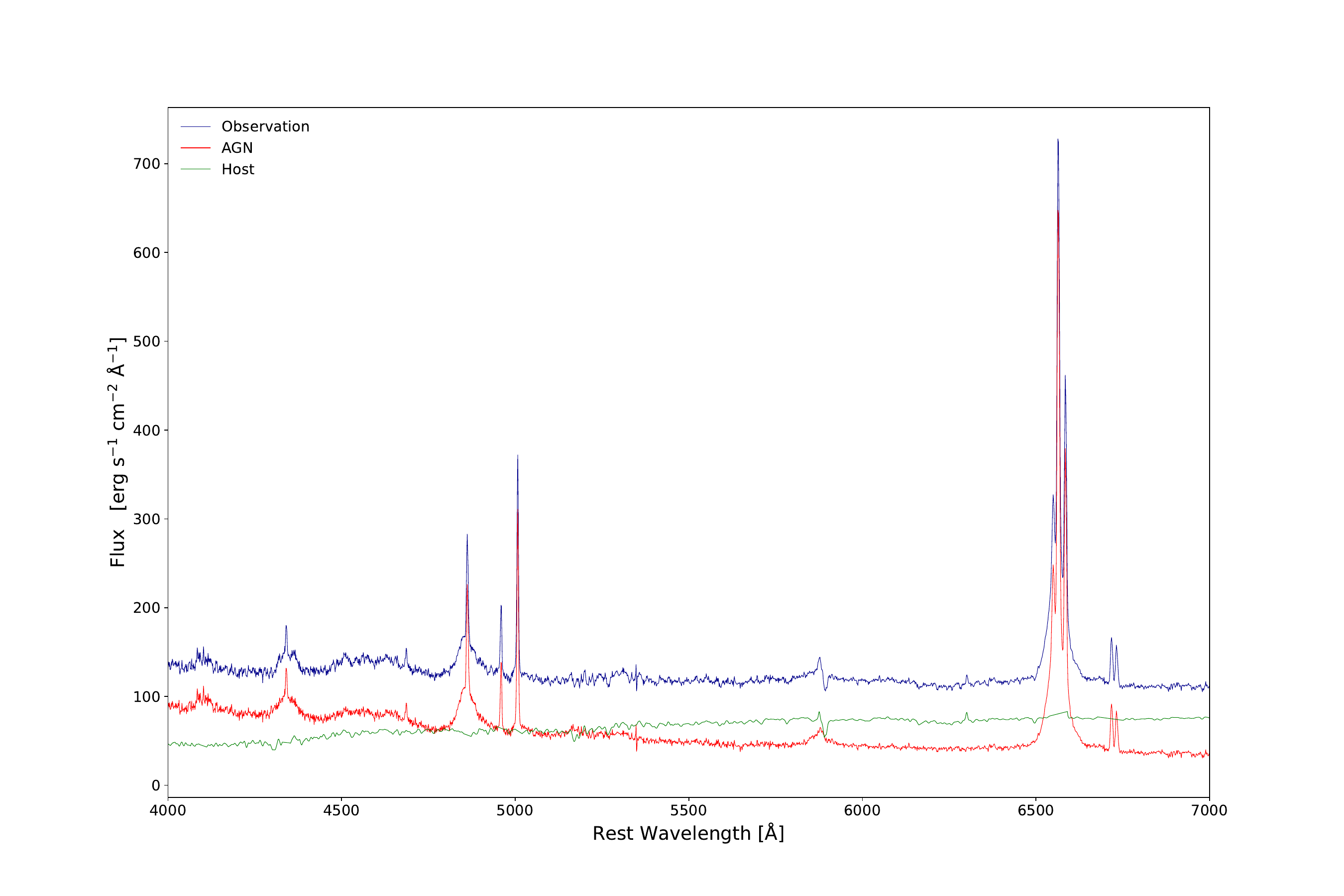}
\caption{PCA decomposition of the whole spectra of spec-1726-53137-0634 {performed with the {\sc fantasy} code}.}
\label{fig:fit28Apr4}
\end{figure*}

\begin{figure*}[h!]
\centering
\includegraphics[width=0.9\textwidth]{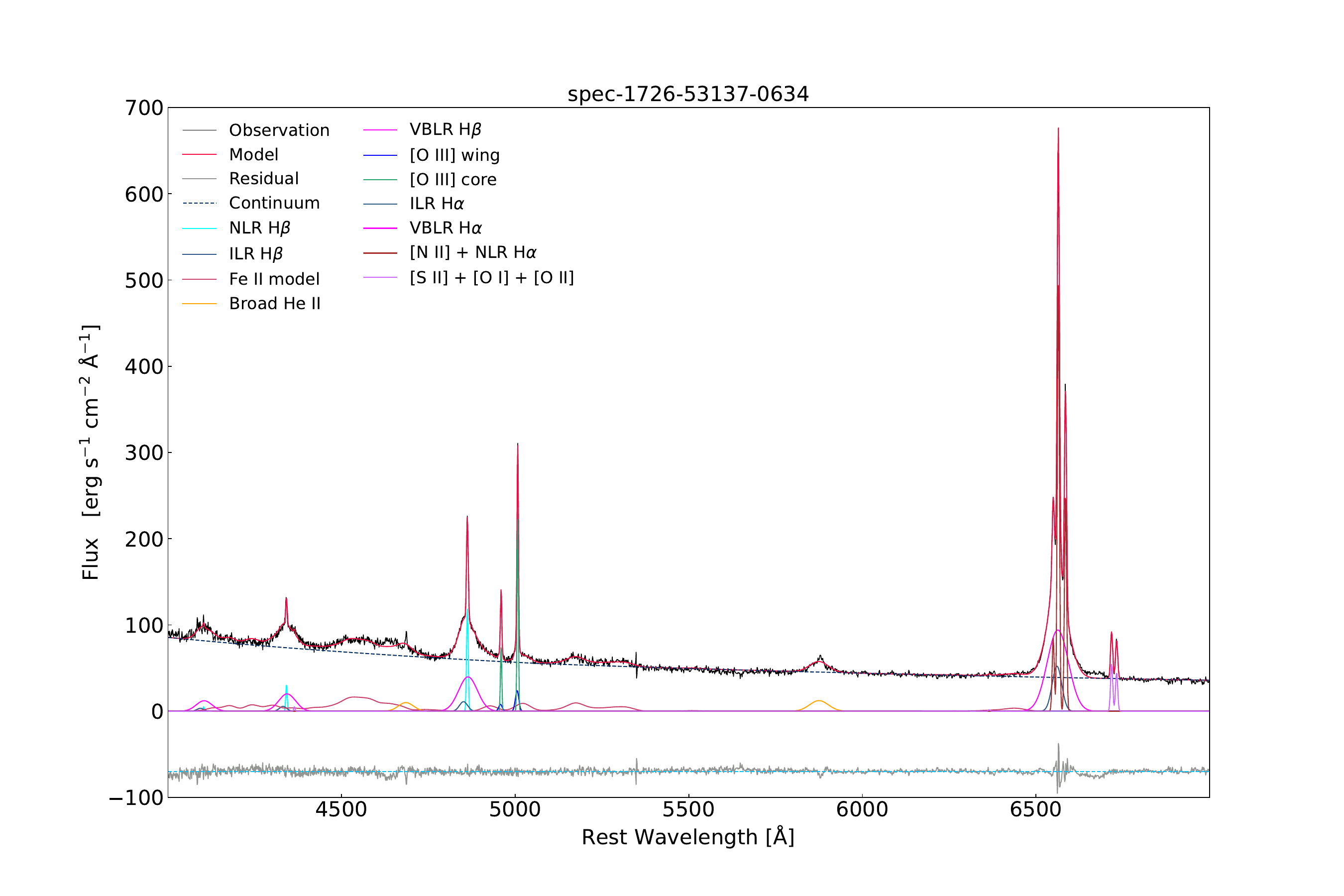}
\caption{Fit to the whole spectra for the source spec-1726-53137-0634 {performed with the {\sc fantasy} code. The fitted components include the NLR, intermediate-line region (ILR), and very broad-line region (VBLR) emission}.}
\label{fig:fit28Apr1}
\end{figure*}

\begin{figure*}[h!]
\centering
\includegraphics[width=0.88\textwidth]{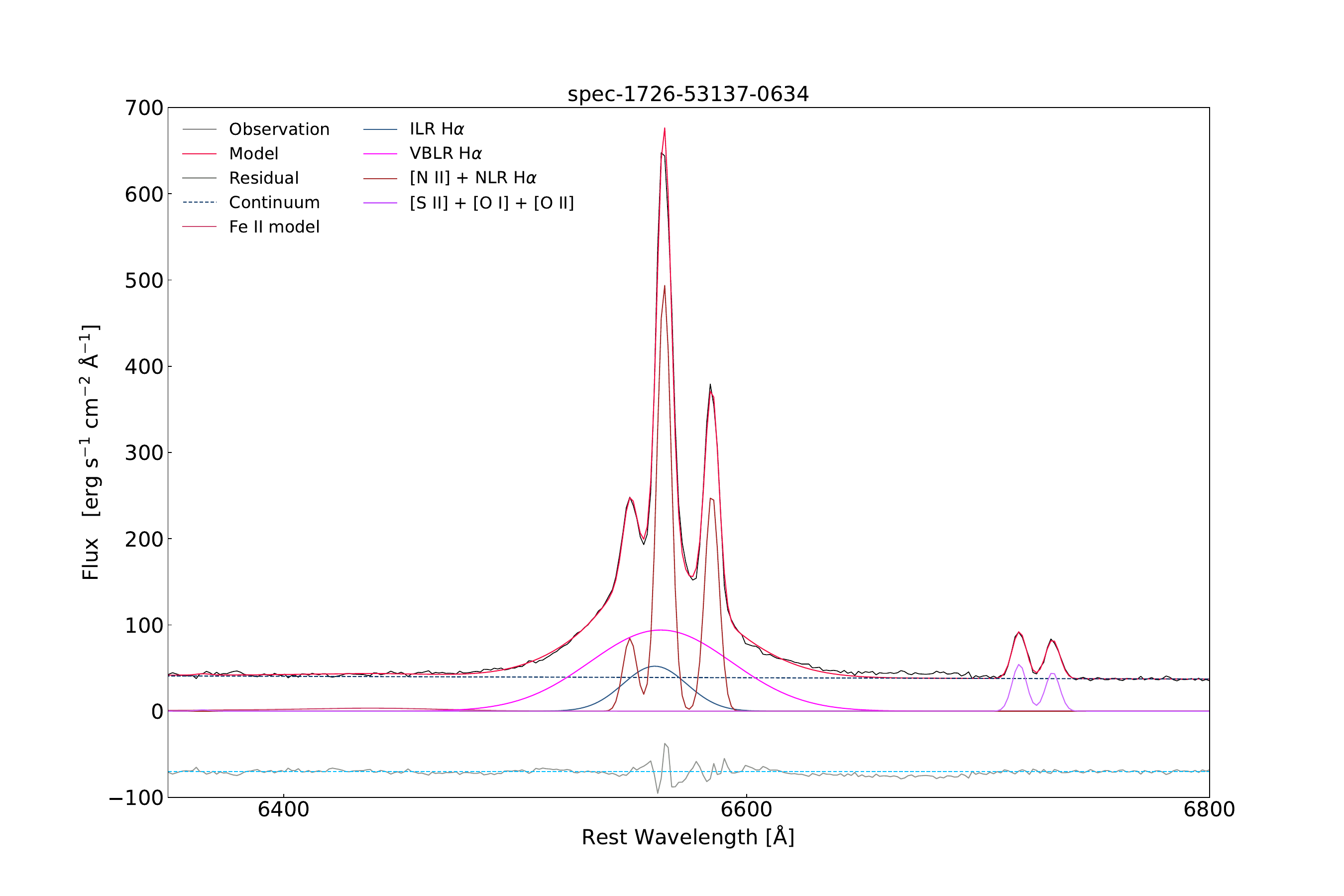}
\caption{Fit to the H$\alpha$ region for source spec-1726-53137-0634 {performed with the {\sc fantasy} code. The fitted components include the NLR, intermediate-line region (ILR), and very broad-line region (VBLR) emission}.}
\label{fig:fit28Apr2}
\end{figure*}

\begin{figure*}[h!]
\centering
\includegraphics[width=0.88\textwidth]{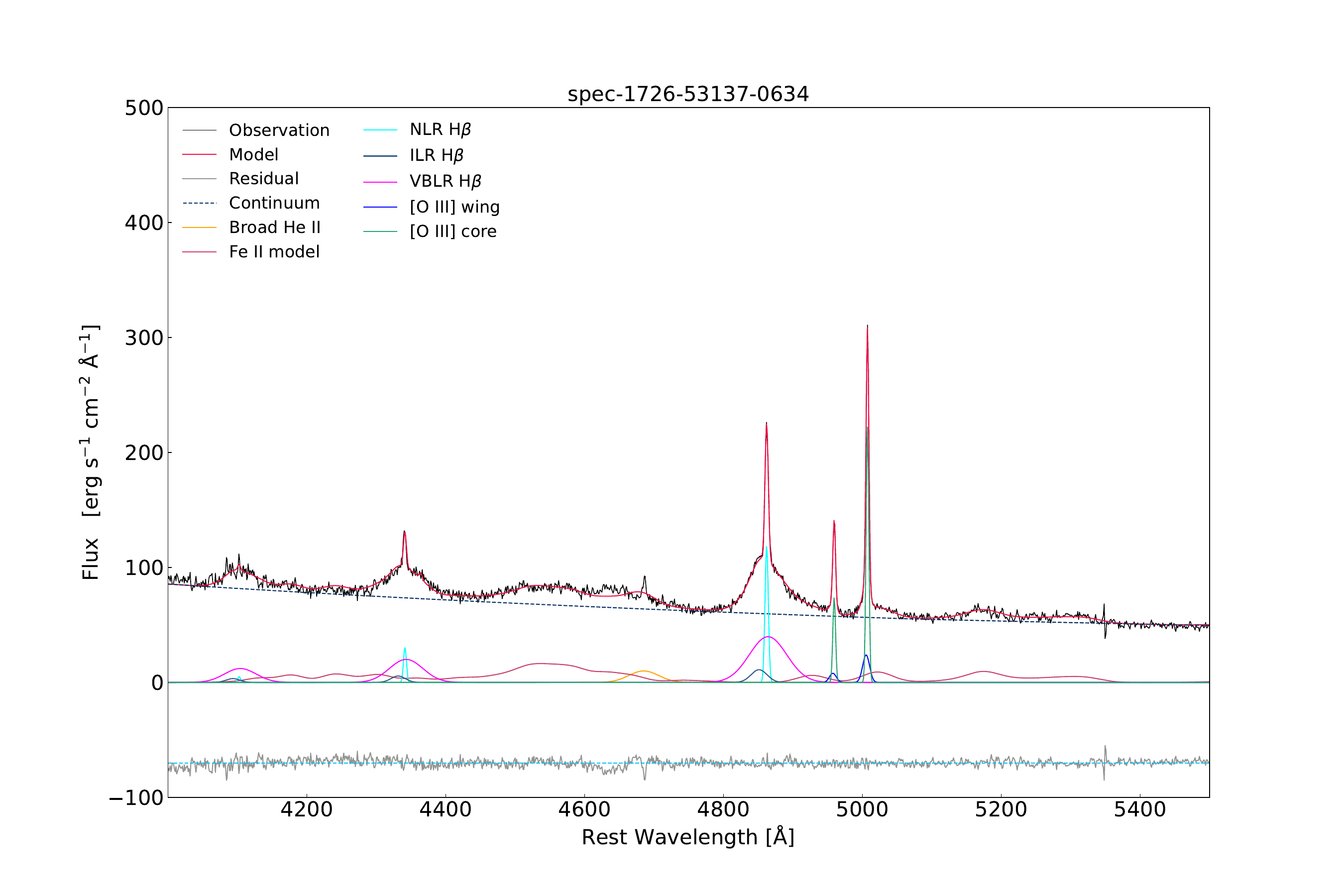}
\caption{Fit to the H$\beta$ region for source spec-1726-53137-0634 {performed with the {\sc fantasy} code. The fitted components include the NLR, intermediate-line region (ILR), and very broad-line region (VBLR) emission}.}
\label{fig:fit28Apr3}
\end{figure*}

An example fit of a single spectrum obtained with the {\sc fantasy} code is presented here. The PCA is displayed: the observed spectrum is decomposed into AGN and host-galaxy components (Fig.~\ref{fig:fit28Apr4}). The final fit of the full spectrum is shown in Fig.~\ref{fig:fit28Apr1}, while the detailed fittings of the H$\alpha$ and H$\beta$ regions are separately illustrated in Figs.~\ref{fig:fit28Apr2} and~\ref{fig:fit28Apr3}, respectively.

\section{Additional relations} \label{sec:wr2}

\begin{figure*}
\centering
\includegraphics[width=150mm]{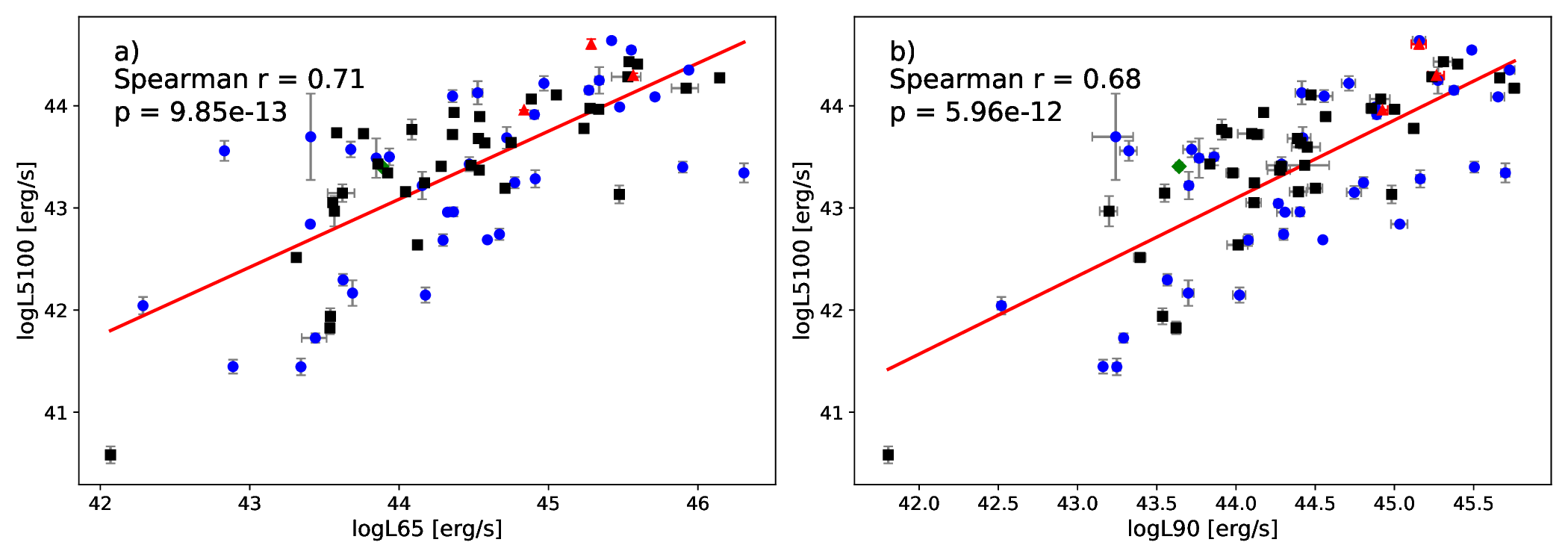}
\includegraphics[width=150mm]{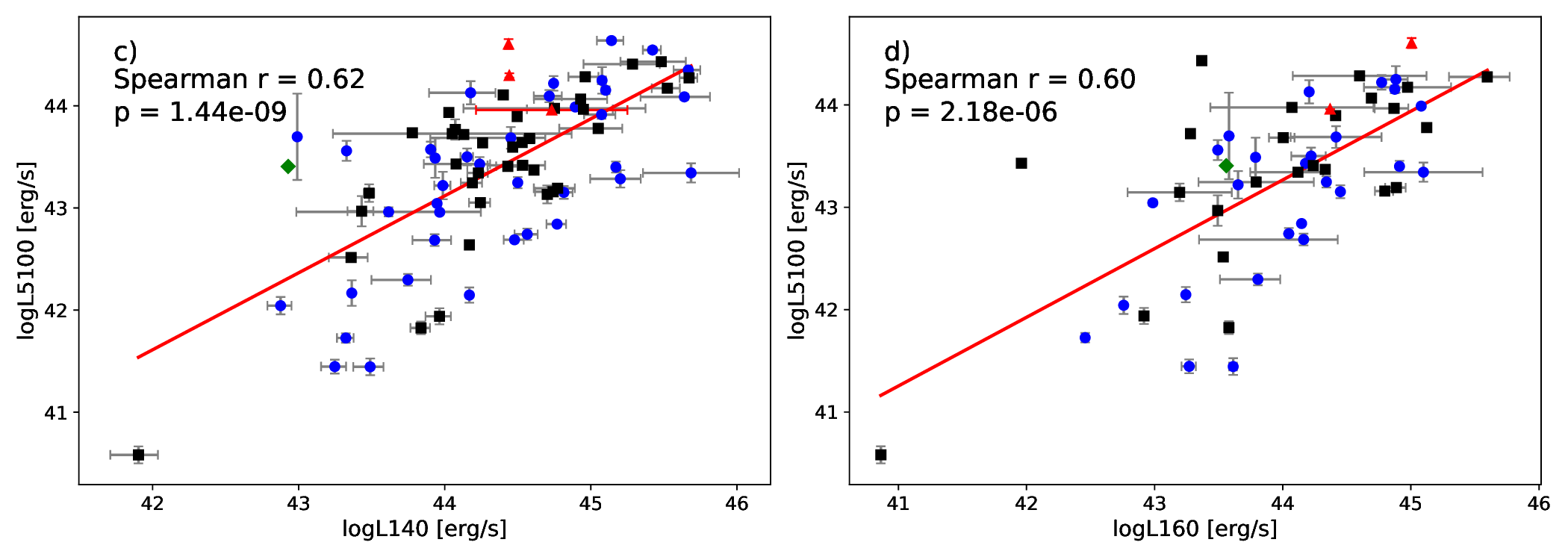}
\caption{Optical luminosity, L5100 vs FIR luminosities at 65, 90, 140 and 160 $\mu$m. \label{fig:owt}}
\end{figure*}

The relations between L5100 and MIR luminosities for the data sample analysed in this study are presented in Fig.~\ref{fig:owt}. 

\begin{figure*}
\centering
\includegraphics[width=182mm]{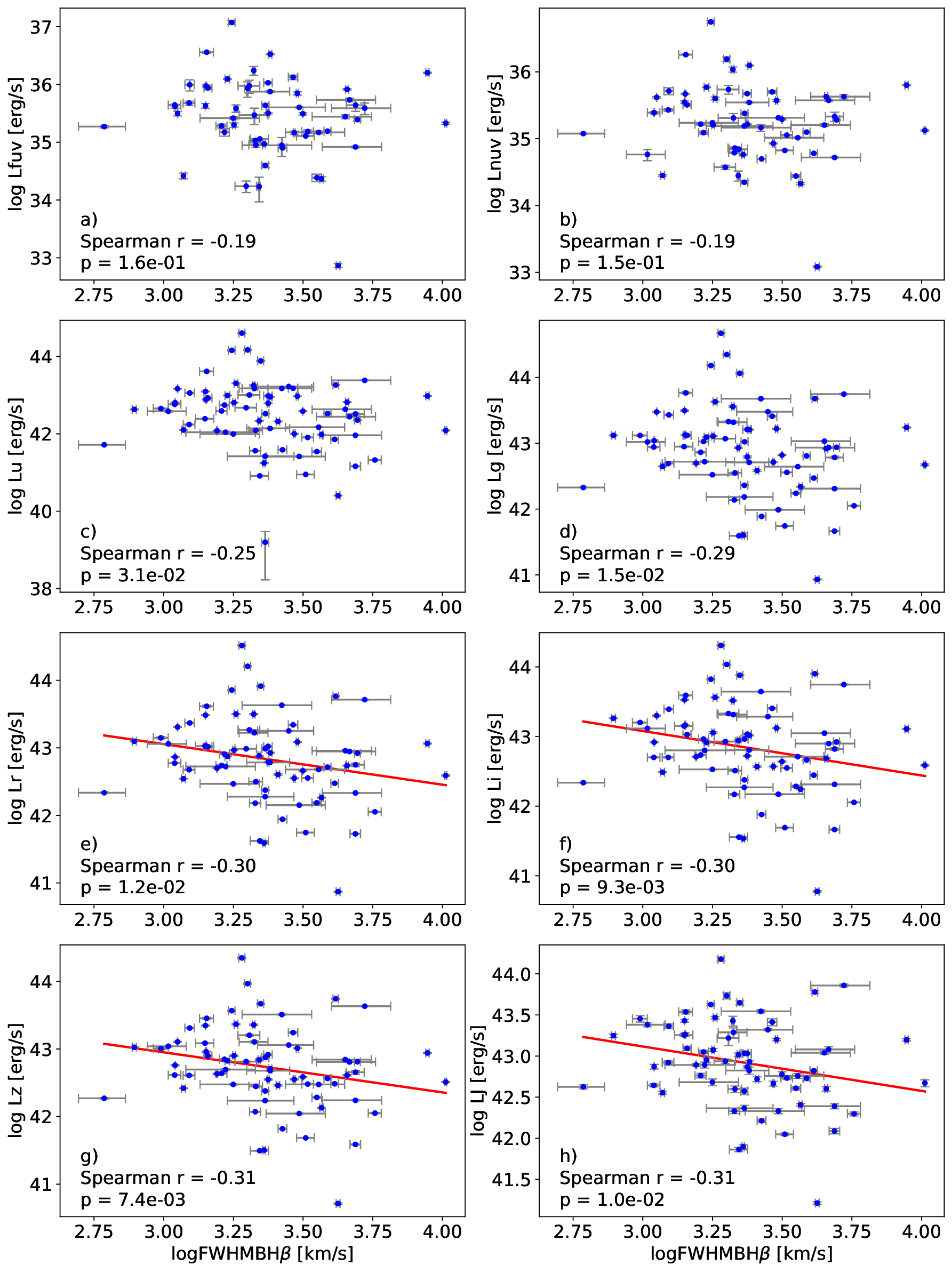}
\caption{FWHMBH$\beta$ vs various UV, optical and NIR luminosities. \label{fig:mul1}}
\end{figure*}

\begin{figure*}
\centering
\includegraphics[width=182mm]{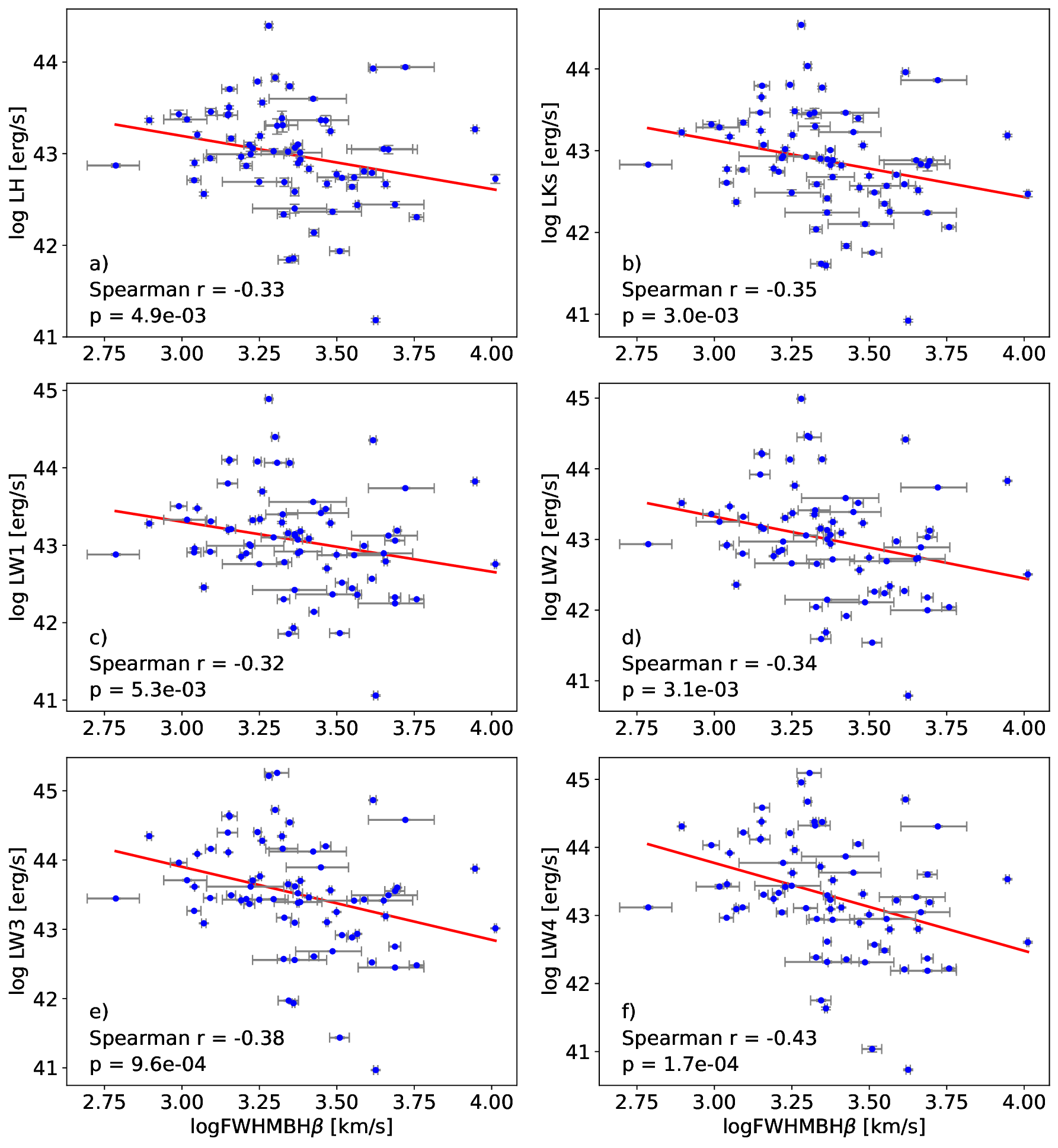}
\caption{FWHMBH$\beta$ vs NIR and MIR luminosities. \label{fig:mul2}}
\end{figure*}

In Figs.~\ref{fig:mul1} and \ref{fig:mul2}, the relations between FWHMBH$\beta$ and luminosities across several bands (FUV, NUV, optical, NIR, and MIR) are examined. It can be observed that certain optical and all IR luminosities display significant anti-correlations with FWHMBH$\beta$. These trends resemble those found between FWHMBH$\beta$ and FIR luminosities (see Fig.~\ref{fig:nize}c,d, Sect.~\ref{sec:wr1}).

\section{M$_{\rm BH}$-L$_{\rm FIR}$ comparison} \label{app:dod}

\begin{figure*}[t]
  \centering
  \includegraphics[width=\textwidth]{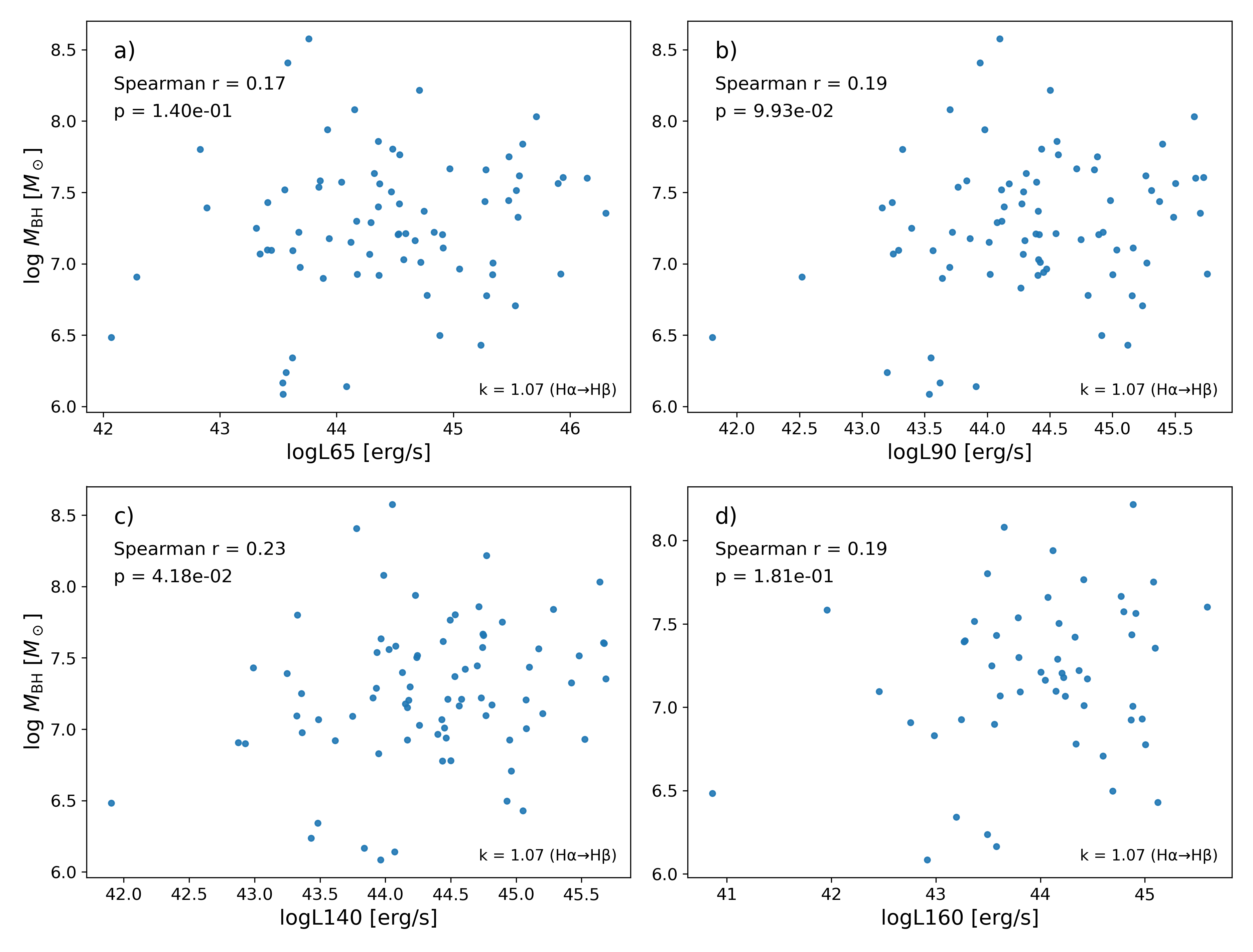}
  \caption{M$_{\rm BH}$ against luminosities at 65, 90, 140, and 160\,$\mu$m, respectively. Each panel reports Spearman $r$ and $p$ value.}
  \label{fig:bhbq}
\end{figure*}

For a consistency check against the Liu19 measurements (Fig.~\ref{fig:mass}), Fig.~\ref{fig:bhbq} shows the scatter between M$_{\rm BH}$ estimates using data from {\sc fantasy} calculation, and FIR luminosities at 65, 90, 140, and 160\,$\mu$m (none of these relations are significant). We estimate M$_{\rm BH}$ with the single-epoch H$\beta$ virial calibration of \citet[][VP06]{VestergaardPeterson2006}, which uses the 5100\,\AA\ continuum luminosity and the H$\beta$ line width:
\[
\log\!\left(\frac{M_{\rm BH}}{M_\odot}\right)
= 6.91
+ 0.50\,\log\!\left(\frac{L_{5100}}{10^{44}\,{\rm erg\,s^{-1}}}\right)
+ 2\,\log\!\left(\frac{{\rm FWHMBH}\beta}{10^{3}\,{\rm km\,s^{-1}}}\right).
\]
When FWHMBH$\beta$ is unavailable, we substitute it using the empirical relation ${\rm FWHMBH\beta} \simeq k\,{\rm FWHMBH\alpha}$, with $k=1.07$ \citep[e.g.][]{Greene2005}.

\section{Comparison of regression methods} \label{app:regression}
\label{sec:appendix_regression}

To assess the robustness of the derived correlations, we compared the results obtained using two different regression approaches. The first method is the standard ordinary least-squares regression implemented via \texttt{scipy.stats.linregress}, which minimises residuals in the dependent variable.

As an alternative estimate, we also computed the orthogonal regression, which treats both variables symmetrically by minimising the orthogonal distances to the regression line. This approach follows the statistical framework described by \citet{FeigelsonBabu2012}, and is commonly adopted in astronomical applications when both variables exhibit significant scatter.

The resulting regression slopes for the log--log relations are presented in Table~\ref{tab:slo}. In each case, the two methods were applied to the same set of data points, ensuring that any differences in the derived slopes are not driven by sample selection effects. In each cell, we list the slope $b$ and its uncertainty $\sigma_b$ for the relation $y = a + bx$, where $x$ corresponds to the luminosity and $y$ to the parameter indicated in the column header.

We find that for line-width parameters (e.g. FWHM), the slopes derived from the two methods are generally consistent within the uncertainties, indicating that the inferred trends are robust with respect to the choice of regression technique. For the black hole mass, the slopes obtained with the two methods are also broadly consistent, although some systematic differences remain, suggesting a moderate sensitivity to the adopted regression method.

Overall, the comparison demonstrates that, for the log--log relations considered here, the main conclusions are not strongly affected by the choice of regression approach, while the orthogonal method tends to yield slightly steeper slopes, as expected when both variables exhibit intrinsic scatter.

\begin{table}
\centering
\scriptsize
\setlength{\tabcolsep}{3pt}
\renewcommand{\arraystretch}{0.92}
\caption{Comparison of regression slopes for the log--log relations, using standard linear regression (LR) and orthogonal regression (OR).}
\label{tab:slo}
\begin{tabular}{llccc}
\hline
Lum. & Met. & FWHMBH$\beta$ & FWHM[O\,III]5007 & logM$_{{\rm BH}}$ \\
\hline
L65 & LR & $-0.11\pm0.03$ & $0.07\pm0.02$ & $0.15\pm0.06$ \\
 & OR & $-0.12\pm0.04$ & $0.08\pm0.02$ & $0.21\pm0.10$ \\
\hline
L90 & LR & $-0.11\pm0.04$ & $0.09\pm0.02$ & $0.18\pm0.07$ \\
 & OR & $-0.12\pm0.04$ & $0.10\pm0.03$ & $0.29\pm0.12$ \\
\hline
L140 & LR & $-0.08\pm0.04$ & $0.10\pm0.03$ & $0.20\pm0.08$ \\
 & OR & $-0.09\pm0.05$ & $0.10\pm0.03$ & $0.35\pm0.13$ \\
\hline
L160 & LR & $-0.06\pm0.04$ & $0.10\pm0.03$ & $0.22\pm0.07$ \\
 & OR & $-0.07\pm0.03$ & $0.11\pm0.03$ & $0.28\pm0.11$ \\
\hline
L5100 & LR & $-0.05\pm0.03$ & $0.08\pm0.02$ & $0.35\pm0.06$ \\
 & OR & $-0.05\pm0.03$ & $0.08\pm0.03$ & $0.48\pm0.10$ \\
\hline
Lpeak & LR & $-0.00\pm0.00$ & $0.00\pm0.00$ & $0.01\pm0.00$ \\
 & OR & $-0.00\pm0.00$ & $0.00\pm0.00$ & $0.01\pm0.01$ \\
\hline
LFUV & LR & $-0.05\pm0.05$ & $0.04\pm0.03$ & $0.32\pm0.09$ \\
 & OR & $-0.06\pm0.04$ & $0.05\pm0.04$ & $0.63\pm0.24$ \\
\hline
LNUV & LR & $-0.07\pm0.05$ & $0.05\pm0.03$ & $0.43\pm0.10$ \\
 & OR & $-0.08\pm0.04$ & $0.05\pm0.05$ & $0.89\pm0.28$ \\
\hline
Lu & LR & $-0.05\pm0.03$ & $0.10\pm0.02$ & $0.28\pm0.06$ \\
 & OR & $-0.06\pm0.03$ & $0.10\pm0.03$ & $0.39\pm0.11$ \\
\hline
Lg & LR & $-0.08\pm0.04$ & $0.15\pm0.03$ & $0.40\pm0.08$ \\
 & OR & $-0.09\pm0.03$ & $0.16\pm0.03$ & $0.67\pm0.14$ \\
\hline
Lr & LR & $-0.09\pm0.04$ & $0.16\pm0.03$ & $0.42\pm0.08$ \\
 & OR & $-0.10\pm0.04$ & $0.17\pm0.03$ & $0.76\pm0.17$ \\
\hline
Li & LR & $-0.09\pm0.04$ & $0.16\pm0.03$ & $0.41\pm0.08$ \\
 & OR & $-0.10\pm0.04$ & $0.17\pm0.03$ & $0.74\pm0.16$ \\
\hline
Lz & LR & $-0.09\pm0.05$ & $0.16\pm0.03$ & $0.42\pm0.09$ \\
 & OR & $-0.11\pm0.04$ & $0.18\pm0.03$ & $0.80\pm0.19$ \\
\hline
LJ & LR & $-0.11\pm0.05$ & $0.19\pm0.03$ & $0.48\pm0.10$ \\
 & OR & $-0.13\pm0.05$ & $0.21\pm0.04$ & $1.06\pm0.28$ \\
\hline
LH & LR & $-0.10\pm0.05$ & $0.19\pm0.03$ & $0.43\pm0.10$ \\
 & OR & $-0.12\pm0.05$ & $0.20\pm0.03$ & $0.96\pm0.27$ \\
\hline
LKs & LR & $-0.10\pm0.04$ & $0.16\pm0.03$ & $0.36\pm0.09$ \\
 & OR & $-0.11\pm0.04$ & $0.17\pm0.03$ & $0.73\pm0.18$ \\
\hline
LW1 & LR & $-0.08\pm0.04$ & $0.13\pm0.03$ & $0.36\pm0.08$ \\
 & OR & $-0.10\pm0.04$ & $0.14\pm0.03$ & $0.66\pm0.16$ \\
\hline
LW2 & LR & $-0.08\pm0.04$ & $0.11\pm0.02$ & $0.28\pm0.07$ \\
 & OR & $-0.09\pm0.03$ & $0.12\pm0.03$ & $0.45\pm0.11$ \\
\hline
LW3 & LR & $-0.09\pm0.03$ & $0.11\pm0.02$ & $0.25\pm0.07$ \\
 & OR & $-0.10\pm0.03$ & $0.12\pm0.03$ & $0.40\pm0.10$ \\
\hline
LW4 & LR & $-0.10\pm0.03$ & $0.11\pm0.02$ & $0.21\pm0.06$ \\
 & OR & $-0.10\pm0.03$ & $0.12\pm0.02$ & $0.31\pm0.09$ \\
\hline
\end{tabular}
\end{table}

\end{appendix}

\end{document}